\documentclass[conference]{IEEEtran}
\usepackage[pdftex]{graphicx}
\graphicspath{{./}{./figures/}}
\DeclareGraphicsExtensions{.pdf,.jpeg,.png}

\usepackage{algorithm}
\usepackage{multirow}
\usepackage{verbatim}
\usepackage{amsmath}
\usepackage{listings}
\usepackage{xcolor}
\usepackage{color, colortbl}
\PassOptionsToPackage{hyphens}{url}
\usepackage[hyphens]{url}
\usepackage{hyperref}
\usepackage{booktabs}
\usepackage{clrscode}
\usepackage{color}
\usepackage{subfigure}
\definecolor{mygreen}{rgb}{0,0.6,0}
\newcommand{\cred}{\textcolor{red}}

\begin{document}

%\title{Algorithm-Directed Persistent Memory \\
%        for High Performance Computing}
        
\title{Algorithm-Directed Crash Consistence in Non-Volatile Memory for HPC}

%\author{\IEEEauthorblockN{xxx}
%\IEEEauthorblockA{xxx \\
%xxx}
%\and
%\IEEEauthorblockN{xxx}
%\IEEEauthorblockA{xxx\\
%xxx\\
%xxx}}
%\and
%\IEEEauthorblockN{James Kirk\\ and Montgomery Scott}
%\IEEEauthorblockA{Starfleet Academy\\
%San Francisco, California 96678--2391\\

\newcommand{\email}[1]{\texttt{\small{#1}}}
\author{
  \begin{minipage}{5.0cm}
    \centering
    Shuo Yang\\
    \email{ys14@mails.tsinghua.edu.cn}
  \end{minipage}
 \begin{minipage}{5.0cm}
    \centering
     Kai Wu$^\dagger$\\
    \email{kwu42@ucmerced.edu}
 \end{minipage}
 \\  \\
  \begin{minipage}{6cm}
    \centering
    Yifan Qiao\\
    \email{qian-yf15@mails.tsinghua.edu.cn}
  \end{minipage}
  \begin{minipage}{5.0cm}
    \centering
    Dong Li$^\dagger$ \\
    \email{dli35@ucmerced.edu}
  \end{minipage} 
 \begin{minipage}{5.0cm}
    \centering
    Jidong Zhai\\
    \email{zhaijidong@tsinghua.edu.cn}
  \end{minipage} \\\\
Tsinghua University \qquad University of California, Merced$^\dagger$
}

\maketitle

%\vspace{-20pt}
\begin{abstract}
Fault tolerance is one of the major design goals for HPC. The emergence of non-volatile memories (NVM) provides a solution to build fault tolerant HPC. Data in NVM-based main memory are not lost when the system crashes because of the non-volatility nature of NVM. However, because of volatile caches, data must be logged and explicitly flushed from caches into NVM to ensure consistence and correctness before crashes, which can cause large runtime overhead. 

In this paper, we introduce an algorithm-based method to establish crash consistence in NVM for HPC applications. We slightly extend application data structures or sparsely flush cache blocks, which introduce ignorable runtime overhead. Such extension or cache flushing allows us to use algorithm knowledge to \textit{reason} data consistence or correct inconsistent data when the application crashes. We demonstrate the effectiveness of our method for three algorithms, including an iterative solver, dense matrix multiplication, and Monte-Carlo simulation. 
Based on comprehensive performance evaluation on a variety of test environments, we demonstrate that our approach 
has very small runtime overhead (at most 8.2\% and less than 3\% in most cases), much smaller than that of traditional checkpoint, while having the same or less recomputation cost.
%\cred{Results show that our approach can reduces 50\% to 80\% runtime overhead compared with the traditional checkpoint and XX\% overhead than a state-of-the-art transaction-based NVM library.} 
%144 words now
\end{abstract}

\section{Introduction}
\label{sec:intro}
%resilience challenge
Fault tolerance is one of the major design goals for HPC. %large-scale HPC systems.
Because of hardware and software faults and errors, HPC applications can
crash or have incorrect computation results during the execution.
\begin{comment}
Building a fault tolerant HPC system will make HPC applications
survive frequent interrupts from faults and errors, and run correctly.
Having such fault tolerance capability is especially important for those scientific applications with long execution time, such as climate simulation, 
because their executions are expected to 
be frequently interrupted in the future HPC systems with a larger number of
system components and possibly higher fault rate.
\end{comment}
The most common strategy to enable fault tolerant HPC is to periodically
store a consistent and correct application state in persistent storage, 
such that there is always a resumable state throughout the application execution.  
Such application state is often characterized as the data values of critical data objects within the application.
If the application crashes or an error is detected during the application execution, the application can go back to the last consistent and correct state, and restart. The application-level checkpoint/restart mechanism is an implementation of such strategy. 

However, there is a problem with the strategy of periodical checkpoint.
If the application state to checkpoint is large, 
the application has to suffer from a large data copy overhead.
This fact is especially pronounced in HPC systems based
on remote storage nodes for checkpoint. 
Although there is a large body of work to reduce the checkpoint overhead,
such as hierarchical checkpoint
to save checkpoint in local compute nodes~\cite{sc10:moody, sc09:dong, sc11:gomez}, incremental checkpoint that only checkpoints modified data to reduce checkpoint size~\cite{isftc94:plank, ics04:agarwal, icpads10:wang, ipdps09:bronevetsky},
and disk-less checkpoint~\cite{tpds98:plank, Lu:2005:SDC:1145057, ppopp17:tang, isftc94:plank}, the checkpoint overhead
remains one of the major scalability challenges for future extreme-scale
HPC systems~\cite{dns16:cappello}.

The emergence of non-volatile memories (NVM), such as phase change memory (PCM) and STT-RAM, provide an alternative solution to build fault tolerant HPC. 
Unlike regular DRAM, those memories are \textit{persistent}, meaning that data are not lost when the system crashes because of the non-volatility nature of NVM. 
Furthermore, short access latency and large memory bandwidth of NVM makes 
the performance of NVM close to that of DRAM. 
In fact, with hardware simulation, the existing work has demonstrated that 
using NVM as the main memory to run sophisticated scientific applications
may not have large performance loss~\cite{nvm_ipdps12}, because of memory level
parallelism and the overlap between computation and memory accesses.
Hence, using NVM as the main memory to build fault tolerant HPC is promising.

However, leveraging the non-volatility of NVM to establish a consistent and correct state, which is called \textit{crash consistence}, 
throughout the application execution in NVM is challenging. 
%in the same way as the checkpoint mechanism does in back-end storage is challenging.
%There are two requirements to establish a consistent state for HPC applications.
Because of volatile hardware caches widely deployed in HPC systems,
%%ensuring a consistent state on NVM %after application crashes 
%%indicates that data must be explicitly flushed from the caches into NVM during the application execution. 
%%Such explicitly-controlled cache flushing is mingled with implicit
%%cache line write-back due to cache coherence protocol and caching algorithms.
there is no guarantee that the application state in NVM is correct and usable by the recovery process to restart applications. Ideally, if the application state in NVM is the same as a one established by the checkpoint mechanism,
then the existing restart mechanism can be seamlessly integrated
into the NVM-based HPC. 
%enforce correct write-orders in NVM; persistent memory

To maintain a consistent and correct state in NVM throughout the application execution,
the most common software-based approaches are redo-log (storing new data updates) or undo-log (storing old data values)~\cite{nv-heaps_asplos11, mnemosyne_asplos11}, 
%see ``ThyNVM: Enabling Software-Transparent Crash Consistency in Persistent Memory Systems'' for further discussion.
and enable a transaction scheme for relatively small workloads (e.g., hash table searching, B-tree searching, and random swap). Those approaches are often based on a programming model with the
support of persistent semantics~\cite{nv-heaps_asplos11, mnemosyne_asplos11, intel_nvm_lib}. 
Those approaches, unfortunately, can impose large overhead, because
they have to log memory update intensively and maintain the corresponding metadata.
Such overhead is especially pronounced when the data objects are
frequently updated by the application. In fact, our preliminary work with CG and dense matrix multiplication based
on a undo-log~\cite{intel_nvm_lib} has 4.3$\times$ and 5.5$\times$ performance loss, respectively.
While such large overhead is tolerable in specific domains (e.g., database) with data persistence prioritized over performance, this overhead is not acceptable
in HPC. To leverage NVM as persistent memory and build a consistent and correct state,
we must introduce a lightweight mechanism with minimum runtime overhead.

In this paper, we introduce a new method to establish a consistent and correct state for critical data objects of HPC applications in NVM. 
The goal is to replace checkpoint for HPC fault tolerance
based on the non-volatility of NVM, 
while introducing ignorable runtime overhead.

Our method is based on the following observation.
Given a relatively small cache size (comparing with the main memory), most of data in an HPC application are not in caches, and should be in consistent state in NVM, because HPC applications are characterized
with a large memory footprint. 
However, how to detect which data in NVM is consistent and can be reused for recomputation is challenging.
The existing logging-based approaches explicitly establish data consistence and correctness with logs, but at the cost of data copy. 
%For an HPC application, most of the data copy may be redundant.
If we can reason the consistent state of data in NVM, then 
we do not need logs, and remove expensive data copy.
%most of data copy can be removed.

Based on the above observation, we propose a novel method %through using algorithm knowledge 
to analyze data consistence and correctness in NVM when the application crashes. 
In particular, instead of frequently tracking and maintaining data consistence and correctness in NVM at runtime,
we slightly extend the application data objects or selectively flush cache blocks, which introduces ignorable runtime overhead. Such application extension or cache flushing allow us to use algorithm knowledge to reason data consistence and correctness when the application crashes.

In essence, we leverage invariant conditions inherent in algorithms, and decide
if the invariant conditions still hold when the crash happens. We study using numerical algorithm knowledge 
to detect consistence and correctness for restart from three perspectives.
First, we leverage the orthogonality relationship between data objects to detect consistence and correctness. 
We use the conjugate algorithm from sparse linear algebra as an example
to study the feasibility of this method.

Second, we leverage the invariant conditions established by the algorithm-based fault tolerance method (ABFT) to detect data consistence and correctness. 
It has been shown that ABFT introduces ignorable runtime overhead by 
slightly embedding extra information (e.g., checksum) into data objects. Using the extra information, we can determine data consistence and correctness when the application crashes, and even correct inconsistent data. 
%where to restart. 
We use an algorithm-based matrix multiplication from dense linear algebra
as an example to study the feasibility of this method.

Third, we study the Monte-Carlo (MC) method. MC
is known for its statistics nature and error tolerance.
In a sense, the inconsistency data is an ``error''. 
Hence MC can restart from the crash without knowing
the consistent state of the critical data objects in NVM.
However, contrary to the common intuition, 
we find that some critical intermediate results in MC
could be lost and have big impact on computation result correctness.
We must ensure the consistence and correctness of those critical intermediate results.
Based on the above algorithm knowledge, we only
flush the data of the critical intermediate results out of caches.
This brings ignorable overhead while ensuring the computation 
correctness when restarting MC.
%when immediately restarting MC from the crash data on NVM.

Algorithm knowledge has been used to address the problems of fault tolerance~\cite{Chen:2013ie, ft_lu_hpdc13, ftfactor_ppopp12, jcs13:wu, abft_ecc:SC13}, performance optimization~\cite{ipdps14:faverge, Lam:asplos91, Williams:sc07}, and energy efficiency~\cite{Dorrance:fpga14, Garcia:lcpc13}. In this paper, we extend the usage of numerical algorithm knowledge into a new territory: by using algorithm-inherent invariant relationship between data objects or algorithm-inherent statistics, we determine consistence and correctness of data objects
without expensive data copy or cache flushing.
%and can restart without introducing expensive data copy.

Our main contributions are summarized as follows.

\begin{itemize}
\item We introduce an algorithm-directed approach to address crash consistence in NVM for HPC. This approach 
has very small runtime overhead (at most 8.2\% and less than 3\% in most cases), much smaller than that of traditional checkpoint, while providing the same or less recomputation cost.  %with the same recomputation cost. 
%reduces 50\% to 80\% runtime overhead compared with checkpoint (hard-drive based) or Intel PMEM (\cred{\textbf{Shuo TODO 70\% to 80\%}}) to maintain data consistence and correctness. 

\item We reveal that with the algorithm-directed approach, the recomputation cost varies because of caching effects. 
%whether the data inconsistence problem is correctable by the algorithm, but 
With a sufficiently large input problem, most of data objects in HPC applications can be consistent and correct in NVM. Hence, the recomputation cost can be small.
%the recomputation cost can be much smaller than the traditional checkpoint.

\item We demonstrate that based on the algorithm-directed approach, leveraging the non-volatility of NVM to enhance or even remove the traditional checkpoint is feasible for future HPC. 
\end{itemize}

\begin{comment}
``while programmers may program directly with persistence primitives, they require
a sophisticated understanding of recovery protocols similar to databases or file system.
Durable transactions provide a much simpler interface to persistent memory and
require few application changes, so we concentrate our evaluation on transaction performance.''

``The cost of durable transaction sis two writes to SCM for every update: xxxx.
Other consistent-update mechanisms may perform better. But, they come at
the cost of increased complexity, such as recovery code to replay logs for append-only updates, garbage collection for memory lost during shadow updates, and explicit fences to order updates.''

``make it persistent by allocating it from a persistent heap and wrapping updates in transactions''

"persist."Memory persistency prescribes the order
of persist operations with respect to one another and loads and
stores, and allows the programmer to reason about guarantees
on the ordering of persists with respect to system failures;
memory persistency is an extension of consistency models
for persistent memory operations. The memory persistency
model relies on the underlying memory consistency model
and volatile memory execution to define persist ordering"
\end{comment}   %1page
\section{Background}
\label{sec:bg}
\begin{comment}
In this section, we review the background information
to discuss the most common checkpoint mechanism
and NVM.

\subsection{Checkpoint for Scientific Applications}
We target on scientific applications in HPC. Most of those applications
are characterized with iterative structures. In particular, 
in those applications, the major scientific simulation
happens in a loop, with each iteration of the loop advancing
one time step of the simulation. The checkpoint happens
after every $n$ time steps. 

The checkpoint can happen at the application level or system level.
At the system level, the whole application state, including heap and stack
data and library data, is checkpointed~\cite{}.
At the application level, the data objects within the application
are selectively checkpointed. 
Those data objects are critical to the application and sufficient to
restart the application. Other data objects that are not checkpointed
can be recomputed based on those critical data objects. 
In a production supercomputer, the application-level checkpoint is
the most common one.
Hence, we focus on the application-level checkpoint and establish a consistent
state on NVM for those critical data objects.

\subsection{Non-Volatile Memory Background}
\end{comment}
Recent progresses on NVM techniques have implemented NVM with different performance characteristics.
%have shown promising characteristics of NVM. 
%There is a large variation in NVM performance for different NVM techniques. 
Some reports have shown that certain NVM techniques (such as ReRAM and STTRAM) 
can achieve very similar latency and bandwidth as DRAM~\cite{NVMDB},
and some NVM techniques (such as PCM) may have less than  an  order  of  magnitude  reduction in  performance (up to 4$\times$ higher latency and 8$\times$ lower bandwidth~\cite{NVMDB, eurosys16:dulloor}). 

Leveraging byte addressability, better scalability, and excellent performance of NVM, using NVM as the main memory is promising. In this NVM usage model, 
NVM may be built as NVDIMM modules, and physically attached to high-speed memory bus and managed by a memory controller~\cite{micro16:chen}.
Furthermore, for those NVM techniques with a performance gap between NVM and DRAM (such as PCM),  the likely deployment of NVM is to build a heterogeneous NVM and DRAM system. In such system, most of the memory is NVM to exploit their low cost
and scalability benefits, and a small fraction of the total memory
is DRAM. A large body of work has explored NVM as the main memory, including those software-based solutions~\cite{eurosys16:dulloor, nas16:giardino, pact15:wei, ics15:guo} and hardware-based solutions~\cite{gpu_pcm_pact13, row_buffer_pcm_iccd12, Ramos:ics11, asplos15:zhang}. 
In this paper, we assume that NVM is used as the main memory, and explore a software-based solution to enable resilient HPC on NVM.

%\textbf{TODO: Explain CLFLUSH instruction.} Have to use CLFLUSH to flush the whole array (no hardware mechanism to track dirty cache lines)

To build a consistent and correct state for critical data objects in NVM, %%and ensure proper recovery, 
the data objects in NVM must be updated with the most recent data in caches. This can be achieved by using cache flushing instructions to flush cache blocks of data objects out of caches. 
Because there is no mechanism to track which cache block is dirty
and whether a specific cache block is in caches, we have to flush
all cache blocks of the data objects, as if those cache blocks are in
caches. Flushing clean cache blocks in caches and flushing cache blocks
not in caches have performance cost at the same order as flushing
dirty cache blocks. Hence, depending on the size of the data objects, flushing cache blocks of the data objects can be expensive.

We use {\fontfamily{qcr}\selectfont CLFLUSH}, the most common cache flush instruction, in this paper.
Other cache flush instructions include 
{\fontfamily{qcr}\selectfont WBINVD}, {\fontfamily{qcr}\selectfont CLFLUSH\_OPT}, and {\fontfamily{qcr}\selectfont CLWB}.
However, {\fontfamily{qcr}\selectfont WBINVD} is a privileged instruction used by operating system (OS); {\fontfamily{qcr}\selectfont CLFLUSH\_OPT} is only available in the most recent Intel processor (skylake) and {\fontfamily{qcr}\selectfont CLWB} is not available in the commercial hardware yet.
Hence we do not use them in the paper. 
But considering them should further improves performance of our proposed approach.

    %1page
\section{Algorithm-Directed Crash Consistence}
In this section, we explain our approach in details for three representative and popular algorithms. We first explain our performance evaluation methodology, commonly used to study the three algorithms.

\begin{comment}
\subsection{General Description}
\label{sec:gen_desc}
In essence, we leverage xxx to detect crash consistence
and decide where to restart.
\end{comment}

\begin{comment}
CG test plan:
For CG tests (Figure 2 to show runtime cost), Use CLASS C to do your tests

For CG tests (Figure 3 to show recomputation cost). 
Test 3.1: The input size is CLASS B and DRAM cache size is 32MB; 
Test 3.2: the input size is CLASS C and DRAM cache size is 32MB
Test 3.3: The input size is CLASS W and DRAM cache size is 32MB; 
Test 3.4: the input size is CLASS S and DRAM cache size is 32MB
Test 3.5: the input size is CLASS D and DRAM cache size is 32MB
\end{comment}

\begin{comment}
XSBench test plan:
 (1) Test runtime overhead with large input size (5678 MB) --> native execution, disk-based checkpoint, NVM-based checkpoint,  our approach (clflush five elements of that macro_xxx array), and pmem (pending).

       (2) Test cache simulator. The cache size will be configured as 32MB.
     2.1 XSbench input problem uses small input 246MB (68 total nuclides, 11303 gridpoints per nuclide)
     2.2 XSbench input problem uses small input 24.6MB. (68 total nuclides, 1130 gridpoints per nuclide)
\end{comment}

\begin{comment}
ABFT test plan:
(1) Runtime overhead test: N=10000 Matrix A about 95MB?.
(2) Recomputation cost: The cache size 32 MB (not 8MB).
Test 2.1: N= 2000;  (N=2000 30MB)
Test 2.2: N= 3000;  (N=3000 68MB)
Test 2.3: N = 4000;  // N=4000 122MB  
Test 2.4: N = 5000;  //N=5000 190MB
N=1500 2000 3000 3500 4000   
  17MB 30MB 69MB 93MB 122MB
\end{comment}

\subsection{Evaluation Methodology}
%%We review our evaluation methodology in this section.
%that is shared between the three algorithms we study in this work in this section. 
To study data consistence between caches and NVM-based main memory when the application crash happens, we develop a ``crash emulator'' that allows us to examine the values of remaining data in caches and main memory. To study application performance in NVM, we use an NVM performance emulator. We explain the crash emulator and test environment as follows.

\textbf{Crash emulator.}
We develop a PIN~\cite{Pin:PLDI05} based crash emulator. In essence, the crash emulator intercepts memory read and write instructions from the application, and emulates a configurable LRU cache. But different from the traditional PIN-based cache emulator, our crash emulator records the most recent data values in caches and main memory. 
The crash emulator also allows the user to specify when to trigger application crash. When a user-specified crash point is triggered, the crash emulator will output the values of data in caches and main memory.

The user can specify the application crash point in two ways.
In the first way, the user can ask the crash emulator to output the data values 
after a specific statement is executed. This is achieved by inserting an API (particularly~\textit{crash\_sim\_output()}) right after the statement in the application.
In the second way, the user can ask the crash emulator to output the data values after a specific number of instructions have been executed. To implement the second way, the crash emulator first profiles the application to collect total number of instructions, and then reports it to the user. The user chooses an instruction number to trigger application crash, and then re-runs the crash emulator. 
The data values will be output by the crash emulator when the user-specified number of instructions is executed.
%which records the most recent data values in caches and main memory, while counting instruction numbers. When the user-specified instruction number is encountered, the data values will be output by the crash emulator. 
%{\color {red} Not very clear about the crash emulator.}

\textbf{Test environment.}
To measure runtime performance, we use a system with two Xeon E5606 (2.13GHz) and 32GB memories. 
To emulate NVM performance, we use Quartz emulator~\cite{middleware15:volos}, a lightweight DRAM-based emulator that allows us to change DRAM bandwidth and latency. 
Quartz has low overhead and good accuracy (with emulation
errors 0.2\% - 9\%)~\cite{middleware15:volos}. More importantly, it allows us to emulate HPC workloads with large data sets within reasonable time.

Since NVM techniques may have inferior performance than DRAM (e.g., 4$\times$ higher latency and 8$\times$ lower bandwidth~\cite{eurosys16:dulloor}), we configure NVM bandwidth as 1/8 of DRAM bandwidth with Quartz. We change memory bandwidth, because our evaluation tests include cache flushing and memory copying, which are sensitive to memory bandwidth.  Since NVM has inferior performance with the lower bandwidth, we introduce a DRAM cache to bridge performance gap between NVM and DRAM. Such heterogeneous NVM/DRAM-based main memory is common~\cite{eurosys16:dulloor, nas16:giardino, hetero_mem_arch, row_buffer_pcm_iccd12, ibm_isca09}. 
The DRAM cache size is 32MB, the same as a recent algorithm-based work for NVM~\cite{hpdc16:wu}. With this heterogeneous NVM/DRAM-based main memory, we must decide data placement between NVM and DRAM. We decide data placement in this memory system based on a recent work~\cite{eurosys16:dulloor}. 

Besides the above NVM emulation based on Quartz, we also study an NVM configuration with the same bandwidth and latency as DRAM. With this configuration, NVM is the same as DRAM. Hence we use an \textit{NVM-only system} without DRAM cache, and do not need Quartz.
Such NVM configuration is based on an optimistic assumption on NVM performance, and eliminates any performance impact of data placement in a heterogeneous NVM/DRAM on our study. 
%To eliminate possible performance impact of data placement for heterogeneous NVM/DRAM and NVM emulation, we also study an NVM configuration with the same bandwidth and latency as DRAM. With this configuration, NVM is the same as DRAM. Hence we use an NVM-only system without DRAM cache, and do not need NVM emulation.

We thoroughly evaluate the performance of our algorithm-based approach with seven test cases.
%We evaluate and compare the performance of seven cases. 
(1) Native execution: the execution without any checkpoint or algorithm-based approach; (2) Checkpoint based on a local hard drive; (3) Checkpoint based on the NVM-only system; (4) Checkpoint based on the heterogeneous NVM/DRAM system; (5) Using the Intel PMEM library~\cite{intel_nvm_lib} on the NVM-only system; (6) Algorithm-based approach on the NVM-only system; (7)  Algorithm-based approach on the heterogeneous NVM/DRAM system.

Among the seven test cases, checkpoint is the most common method to establish a consistent and correct state on non-volatile storage in HPC;
%modern production supercomputer; 
The Intel PMEM library represents the state of the art approach to establish a consistent and correct state in NVM.
We compare our algorithm-based approach with checkpoint and PMEM to study the performance benefit of the algorithm-based approach.

With the memory-based checkpoint (i.e., checkpoint based on the NVM-only system or the heterogeneous NVM/DRAM system), checkpoint is equivalent to
perform data copying plus cache flushing to ensure data consistence between NVM and caches. The cache flushing in the
NVM-only system includes using {\fontfamily{qcr}\selectfont CLFLUSH} instruction
to flush CPU caches; The cache flushing in the heterogeneous NVM/DRAM system includes flushing both CPU caches (using {\fontfamily{qcr}\selectfont CLFLUSH}) and the DRAM cache (using memory copy).

We explain our algorithm-directed crash consistence in details as follows.

\subsection{Algorithm-Directed Crash Consistence for Iterative Method}
\label{sec:cg_pm}

Conjugate Gradient (CG) is one of the most commonly
used  iterative  methods  to  solve  the  sparse  linear  system
$Ax = b$, where the  coefficient  matrix $A$
is  symmetric  positive definite.  
Figure~\ref{fig:cg_pseudocode1} lists the algorithm pseudocode.
In CG, three vectors $p$, $q$, and $z$ can be 
checkpointed for resuming other variables and restarting.
In the rest of this section, we use notation $p^{i}$, $r^{i}$ and $z^{i}$ 
to specify $p$, $r$, and $z$ 
in the iteration $i$
before they are updated in Lines 10, 7, and 5 respectively;
$q^{i}$ specify $q$ in the iteration $i$.

\begin{comment}
before it is written in Line 10 (Figure~\ref{fig:cg_pseudocode1}),  $q$, $r$ before it is written in Line 7 (Figure~\ref{fig:cg_pseudocode1}), and $z$ before it is written in Line 5 (Figure~\ref{fig:cg_pseudocode1}).
\end{comment}

\begin{figure}[!tbp]
\centering
\small
\begin{codebox}
%\Procname{$\proc{Conject Gradiate}$}
\li $r \leftarrow  b - A \cdot x, z \leftarrow 0, p \leftarrow 0, q \leftarrow 0, \rho \leftarrow  {r}^ \mathrm{ T } \cdot r $;
%\State Compute for intialization {r}_{0} {=} {b} - {A}\cdot {x}_{0}, {z}_{0} {=} 0.
%\State {q}_{0} {=} 0, {p}_{0} {=} {r}_{0}, {\rho }_{0} {=} {r}_{0} ^ \mathrm{ T } \cdot {r}_{0} .
\li for $i \gets 1 $ to $ n $  
\li\quad$q \leftarrow A p$
\li\quad$\alpha \leftarrow \rho /({p}^ \mathrm{ T } \cdot q)$
\li\quad$z \leftarrow z + \alpha p$
\li\quad${\rho}_{0} \leftarrow  \rho $
\li\quad$r \leftarrow  r - \alpha p$
\li\quad$\rho \leftarrow  {r}^ \mathrm{ T } \cdot r$
\li\quad$\beta \leftarrow  \rho /{\rho}_{0}$
\li\quad$p \leftarrow  p+ \beta p $
\li\quad Check $r = b - A\cdot z $.
%\End
\li end for
\end{codebox}
\vspace{-10pt}
\caption{Pseudo-code for CG. Capital letters such as $A$ represent matrices; lowercase letters such as $x,y,z$ represent vectors; Greek letters $\alpha, \rho$ represent scalar numbers. }
\label{fig:cg_pseudocode1}
\vspace{-10pt}
\end{figure}

In CG, there are implicit relationships between
multiple data objects, shown in Equations~\ref{eq1:pq} and ~\ref{eq2:rab}.
In particular, Equation~\ref{eq1:pq} shows that at each iteration $i$,
the vectors $p^{(i+1)}$ and $q^{(i)}$ satisfy an orthogonality
relationship.
Equation~\ref{eq2:rab} shows that at each iteration $i$,
the vectors $r^{(i+1)}$, $z^{(i+1)}$, $b$, and the matrix $A$
satisfy an equality relationship.

\begin{equation} \label{eq1:pq}
\small
p^{(i+1)^T} \cdot q^{(i)} = 0
\end{equation}
\begin{equation} \label{eq2:rab}
\small
r^{(i+1)} = b - A\cdot z^{(i+1)}
\end{equation}

\textbf{Algorithm extension.}
Instead of using the checkpoint method (in which at least three arrays  $p$, $q$, and $z$ should be explicitly saved) to achieve the crash consistence, 
%we do not explicitly save any variable or flush caches,
we rely on the existing hardware-based caching mechanism to evict data out of caches and opportunistically
build the crash consistence. 
We extend CG and leverage the above implicit relationships between the data objects
to reason the crash consistency of $p$, $q$, and $z$ in NVM. %and $r$ on NVM. 
This method removes runtime checkpoint and frequent cache flushing, hence improving performance.
\begin{comment}
Once a crash happens, we use an algorithm-based method to detect the state of $p$, $q$, $r$ and $z$ on NVM. %without d CLFLUSH or storing them. 
We leverage the above implicit relationships between the data objects
to reason the crash consistency of $p$, $q$, $r$ and $z$ on NVM. %%once a crash happens, such that we remove the checkpoint operations for the three data objects.
\end{comment}

In particular, if a crash happens at an iteration $i$, 
%without conducting any checkpointing or cache flushing before the cache happens, 
we examine the data values of $p^{(i+1)}$, $q^{(i)}$, $z^{(i+1)}$, and $r^{(i+1)}$ in NVM, and decide if the above implicit relationships are held.
If not, then $p$, $q$, $z$, and $r$ are not consistent and valid, and
we cannot restart from the iteration $i$.
We then check $p^{(i)}$, $q^{(i-1)}$, $z^{(i)}$ and $r^{(i)}$
%to examine if the above implicit relationships are held for
and examine the implicit relationship for the iteration $i-1$.
%If not, we check $p^{(i-1)}$, $q^{(i-2)}$, and $r^{(i-1)}$
%to to examine if the above implicit relationships are held for the iteration $i-2$.
We continue the above process, until we find an iteration
$j$ ($j<i$) where the four data objects satisfy
the above implicit relationship. 
This indicates that $p^{(j+1)}$, $q^{(j)}$, $z^{(j+1)}$, and $r^{(j+1)}$ are consistent and valid. We can restart from the iteration $j$.

To implement the above idea, we need to extend 
the original implementation shown in Figure~\ref{fig:cg_pseudocode1}).
In the figure, $p$, $q$, $r$ and $z$ are one-dimensional arrays overwritten in each iteration.
We add another dimension into the four arrays, such that
each array has the data values of each iteration.
We also flush the cache line containing the iteration number $i$ at the beginning of each iteration.
This makes the iteration number consistent between caches and NVM,
which is helpful for the examination of the data values in NVM after the crash. Note that we only flush one single cache line at every iteration. This brings ignorable performance overhead. Figure~\ref{fig:cg_pseudocode2} shows our extension to the original implementation.

\begin{figure}[htp]
\centering
\vspace{-15pt}
\small
\begin{codebox}
%\Procname{$\proc{Conject Gradiate}$}
\li $r \leftarrow  b - A \cdot x, z \leftarrow 0, p \leftarrow 0, q \leftarrow 0, \rho \leftarrow  {r}^ \mathrm{ T } \cdot r $;
%\State Compute for intialization {r}_{0} {=} {b} - {A}\cdot {x}_{0}, {z}_{0} {=} 0.
%\State {q}_{0} {=} 0, {p}_{0} {=} {r}_{0}, {\rho }_{0} {=} {r}_{0} ^ \mathrm{ T } \cdot {r}_{0} .
\li for $i \gets 1 $ to $ n $  
\li\quad {\color {red} flush the cache line containing $i$}
\li\quad$q[{\color {red} i+1}] \leftarrow A p[{\color {red} i}]$
\li\quad$\alpha \leftarrow \rho /({p}^ \mathrm{ T } \cdot q)$
\li\quad$z[{\color {red} i+1}] \leftarrow z[{\color {red} i}] + \alpha p$
\li\quad${\rho}_{0} \leftarrow  \rho $
\li\quad$r[{\color {red} i+1}] \leftarrow  r[{\color {red} i}] - \alpha p$
\li\quad$\rho \leftarrow  {r}^ \mathrm{ T } \cdot r$
\li\quad$\beta \leftarrow  \rho /{\rho}_{0}$
\li\quad$p[{\color {red} i+1}] \leftarrow  p[{\color {red} i}]+ \beta p $
\li\quad Check $r = b - A\cdot z $.
%\End
\li end for
\end{codebox}
\vspace{-10pt}
\caption{Extending CG to enable algorithm-directed crash consistence. Our extension to CG is highlighted with red color.}
\vspace{-5pt}
\label{fig:cg_pseudocode2}
\end{figure}

\textbf{Performance characterization.}
The above algorithm-based approach does not use checkpoint or frequent cache flushing to explicitly build consistent and valid data state for those critical data objects in NVM. This greatly reduces runtime overhead, as shown in our performance evaluation.
%In essence, our method leverages the existing cache management in hardware (i.e., cache line eviction) to opportunistically establish a consistent and valid data state in NVM.

The success of this approach heavily relies on the memory access patterns in CG. The two-dimensional arrays in the new CG (Figure~\ref{fig:cg_pseudocode2}) have a ``streaming-like'' memory access pattern: the data values generated in any iteration are used in at most two iterations. If the working set size of CG is much larger than the last level cache size, the data values of 
%the three arrays 
$p$, $q$, $r$, and $z$
from the previous iterations before the crash happens have a very good chance to be evicted out of caches and consistent in NVM. In fact, CG is common to be applied on large sparse linear systems that have large data sizes. For those systems, our approach can effectively leverage the existing hardware caching management to evict data out of caches and 
build consistent and valid data states.
In theory, given sufficiently large linear systems, the data values of the four arrays for
a specific iteration $i$ can be evicted out of caches in the iteration $i+1$. If the cache happens at the iteration $i+1$, we
can restart from the iteration $i$, which limits
recomputation cost to only one iteration.
This literally achieves the same recomputation cost as checkpointing at every iteration. 

However, if the linear system to solve by CG is small, the data values of the four
arrays from multiple iterations are in caches and get lost when the crash happens. Depending on how many iterations of the data are lost, our approach could result in larger recomputation cost than the traditional checkpoint.
In the worst case, our approach has to restart from the very beginning of CG. However, for those small systems with data sizes smaller than the last level cache size, the recomputation cost is small, and hence may not be a problem.

\textbf{Performance evaluation.}
%For CG, three data objects are checkpointed, i.e., the arrays $p[]$, $r[]$ and $q$. These three data objects are enough to resume other major data objects during recomputation.
%Firstly, we use our method and simulate the state of cache and memory using our cache simulator. Then 
We evaluate the performance of our approach from two perspectives, recomputation cost and runtime overhead. Ideally, we want to minimize recomputation cost after crashes, and minimize runtime overhead.

\begin{figure}
\centering
\includegraphics[width=0.48\textwidth, height=0.15\textheight]{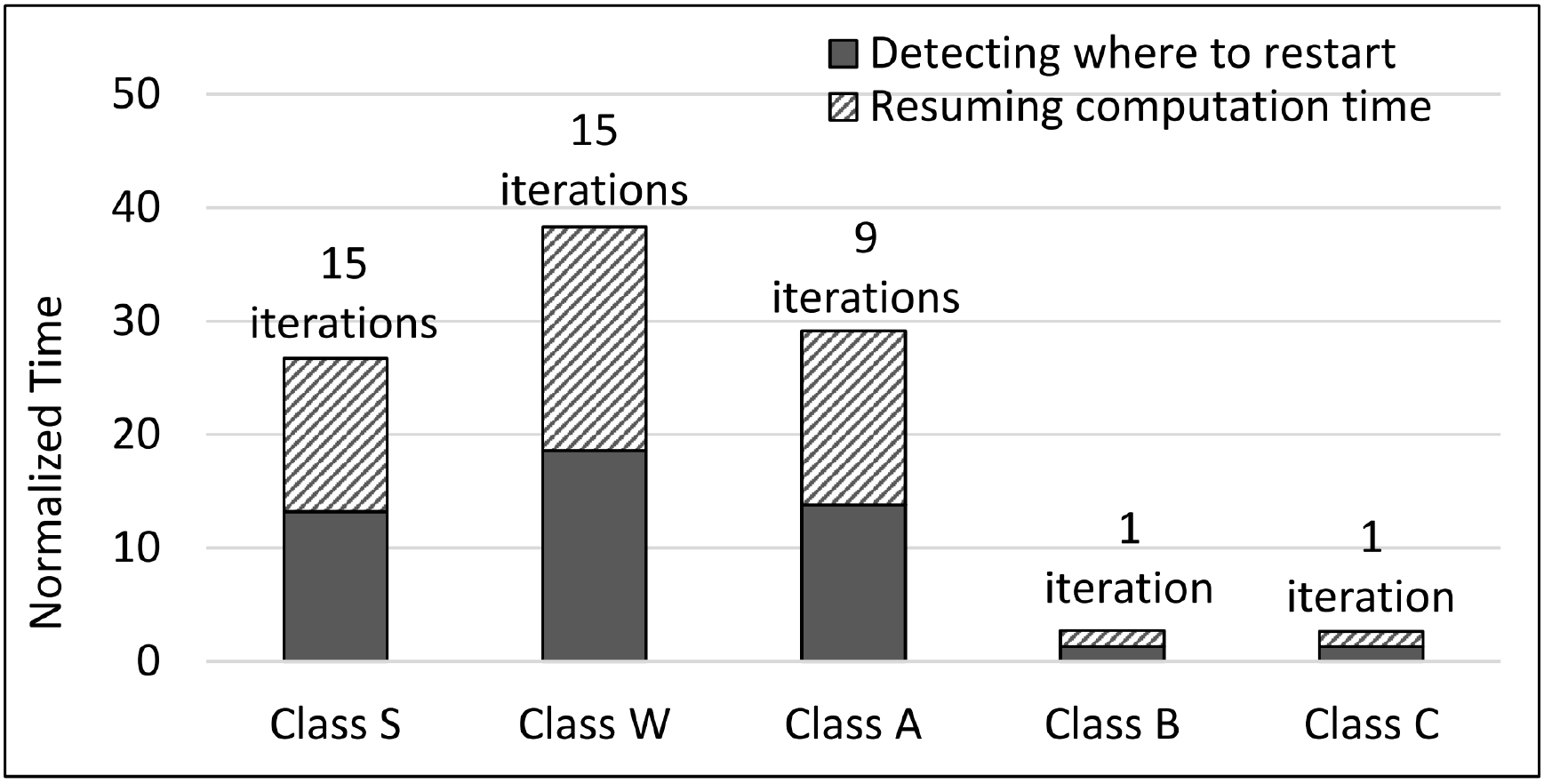}
\vspace{-5pt}
\caption{Recomputation cost (execution time) for CG with our algorithm-based approach. The recomputation cost is normalized by the average execution time of individual iterations of CG. The input problem size is ranked in the increasing order in the $x$ axis.}
\label{fig:cg_recomp_cost}
\vspace{-15pt}
\end{figure}

To measure recomputation cost, we use the crash emulator to trigger a crash at a specific program execution point, particularly Line 10 (Figure~\ref{fig:cg_pseudocode2}) in the 15th iteration of the main loop in NPB CG (one benchmark in NAS parallel benchmark suite~\cite{nas}). Figure~\ref{fig:cg_recomp_cost} shows the performance on the heterogeneous NVM/DRAM system with different input problems of CG. The recomputation time in the figure is broken down, and it includes the time to detect from which iteration CG is resumable (labeled as ``Detecting where to restart'') and the time to resume from the resumable iteration to the crashed iteration (labeled as ``Resuming computation time'' in the figure). 
The recomputation time is normalized by the average execution time of individual iterations of the main loop in CG. 
The number of iterations on the top of each column
is the number of iteration we lose because of the crash.

Figure~\ref{fig:cg_recomp_cost} reveals that the recomputation cost becomes smaller when we use a larger input problem size.
When the input problem size is small (Classes S and W), the recompaution time is relatively large. We lose all of the iterations (15 iterations) when the crash happens. However, when the input problem size is large (Classes B and C), we lose only 1 iteration and the recomputation time is very small. 
This result is aligned with our performance characterization:
in particular, a larger input problem tends to lose smaller computation when a crash happens. 

\begin{comment}
When the input problem is small, the values of the critical data objects in the previous iterations are able to be cached without being flushed to NVM. Hence, when a crash happens, we lose the computation results of previous iterations. However, when the input problem size is bigger, the previous computation results have less opportunities to be cached, because of the streaming access pattern associated with the critical data objects we introduce into CG. This streaming access pattern is characterized with that the array elements from an individual low dimension is accessed only once in an iteration, but not accessed in the other iterations.
\end{comment}

%Paragraph 9: explain that based on the above figure, we can reduce the frequency of doing data copy, because the recompuation is limited to only one iteration in most cases.

\begin{figure}
\centering
\includegraphics[width=0.45\textwidth, height=0.15\textheight]{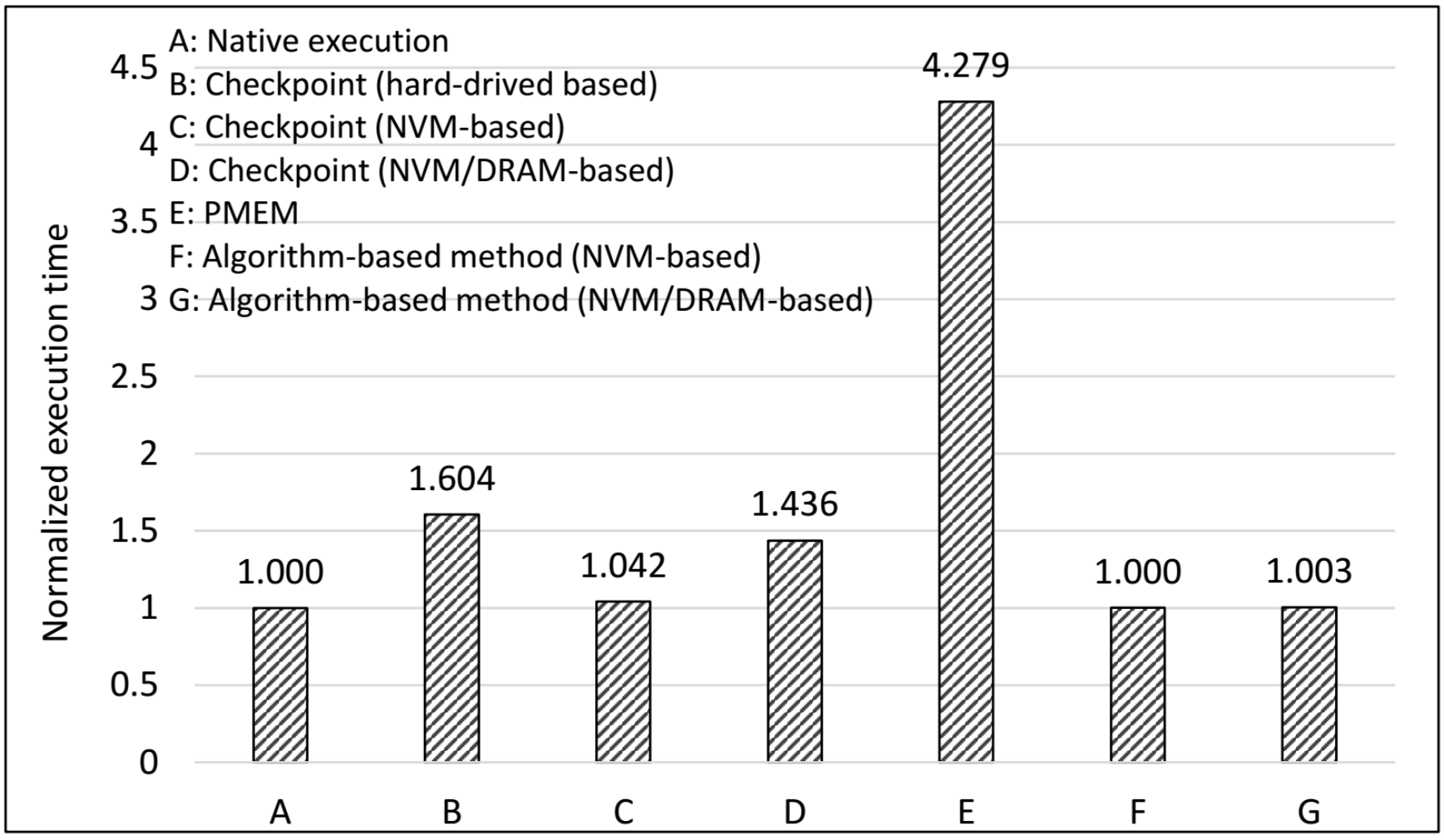}
\vspace{-10pt}
\caption{Runtime performance (execution time) with various mechanisms and our algorithm-based approach to enable crash consistence. Performance is normalized by the performance of native execution without any checkpoint or algorithm extension.} 
%Performance is normalized to the native execution time with neither checkpoint nor our algorithm-based approach.
\label{fig:cg_runtime}
\vspace{-15pt}
\end{figure}

We further compare runtime overhead between traditional checkpoint, Intel PMEM library~\cite{intel_nvm_lib}, and our approach. 
%Checkpoint is the most common method to establish a consistent state on non-volatile storage in modern production supercomputer.
%The Intel PMEM library represents the state of the art approach to establish a consistent state on NVM. We use CLASS C as the input problem.
With Class C as the input problem, recomputation with our approach is limited to one iteration, already shown in Figure~\ref{fig:cg_recomp_cost}.
%To be aligned with this recomputation cost, 
Hence, we make checkpoint at the end of each iteration of the main loop in CG. This frequent checkpoint enables a fair performance comparison, because checkpoint at the end of each iteration results in the same recomputation cost as our algorithm-based approach. 
For the PMEM library, we use its transaction mechanism and enable transactional updates on the three arrays (i.e., $p$, $r$, and $z$). Each iteration of the main loop of CG is a transaction, which makes the recomputation cost with PMEM also limited to one iteration. 
The transaction mechanism in the PMEM library is based on undo log. Figure~\ref{fig:cg_runtime} shows the results.  

Figure~\ref{fig:cg_runtime} reveals that the traditional checkpoint based on hard drive has very large overhead (60.4\%), comparing with the native execution), even if the hard drive is local. Assuming that NVM has equivalent performance as DRAM, NVM-based checkpoint leads to ignorable overhead (4.2\%).  
However, if NVM performance is not as good as DRAM, frequent checkpoint causes large overhead (43.6\%).
Our further study reveals that 51.9\% of the overhead comes from data copying and 48.1\% comes from cache flushing (including DRAM cache flushing).
The PMEM library also causes large overhead (329\%), because of frequent and expensive data logging. 
Our approach, on the other hand, has ignorable runtime overhead (less than 3\%, no matter whether on NVM-only or NVM/DRAM system) because our approach does not come with any extra data copying and only flushes one single cache line per iteration.

\textbf{Conclusions.} Our evaluation results demonstrate that CG with our algorithm extension achieves superior runtime performance with close to zero runtime overhead. Furthermore, when the input problem size is large enough, the recomputation cost is limited to only one iteration. Our approach performs better than the traditional checkpoint or PMEM library for NVM. 

\begin{comment}
Performance evaluation: (1) Figure 2: runtime overhead; 
Our approach vs. native execution (no checkpoint) vs. checkpoint execution (NVM-based checkpoint happens at the end of each iteration) vs. Intel pmem-based implementation

(2) Test re-computation.
Figure 3: cost with different input problem sizes and cache sizes.
We fixed a point to do the crash tests (e.g., in the middle of the ten iteration).
\end{comment}

% \begin{thebibliography}{0}
% \bibitem{HestnessStiefel}
% M. Hestenes, E. Steifel, "Method of conjugate gradients for solving linear sysems", J. Res. Nat. Bur. Standard., vol. 49, no. 152, pp. 409-436.
% \bibitem{Baileyetc}
% THE NAS PARALLEL BENCHMARKS
% \end{thebibliography}

\subsection{Algorithm-Directed Crash Consistence for Matrix Multiplication}
\label{sec:mm_abft_pm}
Matrix multiplication ($C = A \times B$) is a very common numerical computation. 
Matrix multiplication (MM) has been extended to handle fail-continue errors which have the failed process to continue working when errors occur. 
To detect and correct fail-continue errors, MM 
is extended to add checksum within matrices $A$, $B$, and $C$.  
Such method is based on numerical algorithm (called \textit{algorithm-based fault tolerance or ABFT}) has shown success within 5\% performance loss while detecting and correcting certain number of errors in the matrices.
%We leverage the checksum information to detect crash consistence and correct inconsistent data in the output matrix $C$ in NVM. 

How MM is extended to add checksums in the existing method~\cite{jcs13:wu} is as follows.
The input matrices $A$ and $B$ are encoded into a new form with checksums, shown in Equations~\ref{eq:abft_a} and~\ref{eq:abft_b}.
Equation~\ref{eq:abft_a} defines a column checksum matrix of $A$ (assuming the size of $A$ is $m \times k$), %($m \times k$),
denoted by $A_c$, where the vector $v$ is a checksum vector. $v$ is typical set with all elements as 1. 
With $v$, the last row of $A_c$ ($a_{m+1,j}$, $1 \leq j \leq k$) is also shown in Equation~\ref{eq:abft_a}.
Equation~\ref{eq:abft_b} defines a row checksum matrix of matrix $B$ (assuming the size of $B$ is $k \times n$), where the vector $w$ is a checksum vector. $w$ is typically set with all elements as 1.  With $w$, the last column of $B_r$ ($b_{j,n+1}$, $1 \leq j \leq k$) is also shown in Equation~\ref{eq:abft_b}.

\begin{equation}
\label{eq:abft_a}
\small
A_{c}=
\left(\begin{array}{cccc}   
    A \\      
    v_{c}^{\mathrm{ T }}A   
\end{array}\right),
\quad a_{m+1,j}= \sum_{i=1}^{m} a_{i,j}
\end{equation}

\begin{equation}
\label{eq:abft_b}
\small
B_{r}=
\left(\begin{array}{cccc}   
    B & Bv_{r} \\  
\end{array}\right),
\quad b_{j,n+1}= \sum_{j=1}^{n} b_{i,j}
\end{equation}

%In the ABFT, instead of getting $C$ from $AB$, we get the checksum version matrix $C^{f}$ from $C^{f}=A_{c} \times B_{r}$. We just add one row and one colummn which will not add much runtime overhead.
With the encoded $A$ and $B$, instead of computing $C = A \times B$,
we compute their checksum versions, shown in Equation~\ref{eq:abft_c}.
$C^f$ is the result matrix with checksums. 
In particular, in $C^f$, the summations of each row of $C$ are stored in the 
extra column of $C^f$, and the summation of each column of $C$ are stored in the extra row of $C^f$, shown in Equation~\ref{eq:abft_c_elements}.
If one element of $C$ is corrupted, using the checksum relationship shown in Equation~\ref{eq:abft_c_elements}, we can detect and correct errors.

\begin{equation}
\label{eq:abft_c}
\small
C^{f}= A_c \times B_r = 
\left(\begin{array}{cccc}   
    AB & ABv_{r} \\      
    v_{c}^{\mathrm{ T }}AB & v_{c}^{\mathrm{ T }}ABv_{r}   
\end{array}\right)
\end{equation}

%By adding more number, we add more data redundancy and dependence relationship which can be exploited to detect the error.
%Base on the above relationship in matrix $A$ and matrix $B$.
%In the matrix $C^{f}$, there exist.

\begin{equation}
\label{eq:abft_c_elements}
\small
c_{m+1,j}= \sum_{i=1}^{m} c_{i,j},\quad c_{i,n+1}= \sum_{j=1}^{n} c_{i,j}
\end{equation}

The above ABFT is commonly implemented based on Figure~\ref{fig:abft_pseudocode1}. This implementation detects errors at every iteration of the loop (Line 2 in Figure~\ref{fig:abft_pseudocode1}) and makes better use of cache system based on a rank $k$ update (Line 1 in Figure~\ref{fig:abft_pseudocode1})~\footnote{For brevity, we assume $A$ and $B$ are $n$ by $n$ square matrix, and $(n+1)$ is divisible by $k$.}.
Within each iteration of the loop, this implementation does a submatrix multiplication and accumulates the result into $C^f$. 
%is a blocking algorithm to make better use of caches. 
Our following discussion is based on this implementation.

\begin{figure}
\centering
\small
\begin{codebox}
%\Procname{$\proc{ABFT matrix multiplication}$}
\li \textbf{for} $s=1$; $s\leq n+1$ ; $s \leftarrow s+k$
\li \quad Verify the checksum relationship of $C^f$
\li \quad $C^f \leftarrow C^f + A_{c}(1:n+1, s:s+k-1) \times $ \\    
 \quad \quad \quad \quad \quad \quad $B_{r}(s:s+k-1, 1:n+1)$
\li \textbf{end for}
%\li \textbf{if} $s \leq k$ else
%\li \quad $C^f \leftarrow C^f + A_{c}(1:n+1, s:n+1)$ \\ 
% \quad \quad \quad \quad \quad \quad  $\times B_{r}(s:n+1, 1:n+1) $
%\li \textbf{end if}
\end{codebox}
\vspace{-10pt}
\caption{A practical implementation of ABFT for matrix multiplication.}
\label{fig:abft_pseudocode1}
\vspace{-15pt}
\end{figure}

\begin{comment}
\begin{lstlisting}[language=c++,caption=Algorithm based fault tolerance for matrix multiplication, label=code:collective_write, linewidth=9cm]
for (s=1; s<=abs(k/rank)*rank; s=s+rank) {
   Verify the checksum relationship of $C^f$
   $C^f = C^f + A_{c}(1:n+1, s:s+k-1)\times B_{r}(s:s+k-1, 1:n+1) $
}

if s <= k else
   $C^f \leftarrow C^f + A_{c}(1:n+1, s:n+1) \times B_{r}(s:n+1, 1:n+1) $
\end{lstlisting}
\end{comment}

\textbf{Algorithm extension.} 
We leverage the checksum information to detect crash consistence and correct inconsistent data in the output matrix $C$ in NVM. 
%To detect crash consistence of $C$ in NVM based on matrix checksums in ABFT,
To do so, 
a naive idea is to flush checksums at the end of each iteration of the loop in Figure~\ref{fig:abft_pseudocode1}.
Then, when a crash happens, we check checksums to detect the validness
of matrix rows and columns in $C$.
However, the above idea does not work for the following two reasons.

First, we cannot detect consistence in the middle of an iteration based on the checksums.
%The checksums are overwritten in each iteration.
Checksums are only useful to detect consistence at the beginning of each iteration.
%Although the checksums become consistent at the end of each iteration based on the naive idea,  they are partially updated in the middle of each iteration and not useful to dde. 
In the middle of each iteration, the checksum row and column in $C$ may be partially updated by computation, and Equation~\ref{eq:abft_c_elements} is not held.
%hence cannot be used to detect consistence and decide where to restart when the crash happens.
Second, the matrix $C^f$ is completely overwritten in each iteration,
hence restart will be difficult. Even if we find inconsistence in $C^f$ (i.e., some elements of $C_f$ come from the iteration $i$ while other elements come from the iteration $j$ ($j \ne i$)),
we do not have a consistent copy of $C^f$ (i.e., all elements of $C^f$ come from the same iteration) to restart.
Checksums may be able to correct some of inconsistent elements based on Equation~\ref{eq:abft_c_elements} at the beginning of the iteration,
but such checksums can only correct
limited number of inconsistent elements.
%having multiple inconsistent elements in a row or in a column  can be beyond the correction capability of checksums.
%Furthermore, %since $C_f$ is completely overwritten, 
%multiple elements in a row or in a column can be corrupted and not correctable by the checksums.

%\textbf{Algorithm analysis.}

To address the above problems, we extend the above naive idea and introduce
a new algorithm shown in Figure~\ref{fig:abft_pseudocode2}. 
The new algorithm decomposes the loop in the original ABFT into two loops.
One loop performs the submatrix multiplication, and the other loop performs
the addition of the submatrix multiplication results.

In the first loop, the submatrices are still the same as those in the original ABFT with embedded checksums. But different from the original ABFT, the first loop
saves submatrix multiplication results into temporal matrices $C_s^{temp}$ ($s=1,..., (n+1)/k$) with row and column checksums, and flushes those checksums to make sure they are consistent (Line 5 in Figure~\ref{fig:abft_pseudocode2}). If a crash happens in the first loop, then using those checksums we can
detect which temporal matrix ($C_s^{temp}$) is inconsistent in NVM.
Any inconsistent and uncorrectable temporal matrix by checksums will be recomputed. 

In the second loop, we perform the addition (matrix addition) of $C_s^{temp}$ ($s=1,..., (n+1)/k$),
and save the result with row checksum embedded in a temporal matrix $C_{temp}$.
Note that we perform matrix addition rows by rows. Hence the row checksums, once established and consistent in NVM (Line 13 in Figure~\ref{fig:abft_pseudocode2}), will not be overwritten. 
If the crash happens in the second loop, then the row checksums in $C_{temp}$
can decide which rows are not consistent and should be recalculated. 

\begin{figure}[btp]
\centering
\small
\vspace{-15pt}
\begin{codebox}
%\Procname{$\proc{ABFT matrix multiplication}$}
%\li \textbf{for} $s=1$; $s\leq n+1 $ ; $s \leftarrow s+k$
\li //submatrix multiplication $C_s^{temp}(.) = A_c(.) \times B_r(.)$
\li \textbf{for} $s=1$; $s\leq (n+1)/k $ ; $s \leftarrow s+1$
\li \quad $C_s^{temp} \leftarrow A_c(1:n+1, \; (s-1) \times k+1:s\times k) \;\;  \times$ 
\li \qquad \qquad $B_{r}((s-1) \times k+1:s\times k, \; 1:n+1)$
\li \quad flush row and column checksums in $C_s^{temp}$
\li \textbf{end for}
\li  
\li  //submatrix addition 
\li \textbf{for} $i=1$; $i\leq n+1 $ ; $i \leftarrow i+k$
%%\li \quad ${\color {red} C_{i+1} \leftarrow C_{i+1} + C_{i} }$
\li \quad $C_{temp}(i:(i+k-1), \; 1:n+1) \leftarrow$ 
\li \qquad \qquad  $C_{temp}(i:(i+k-1), \; 1:n+1)$ +  
\li \qquad \qquad $\sum_{s=1}^{n/k}C_s^{temp}(i:(i+k-1), \; 1:n+1) $%%\\{\color {red} (not need 11 to 13) }
%%\li \quad {\color {red} flush row checksums and col checksums of  $ C_{i+1}$}
\li \quad flush $k$ rows of row checksums in $C_{temp}$  \\
    \quad (particularly, $C_{temp}(i:(i+k-1), n+1)$)  %%\\ {\color {red} (not need 15) }
\li \textbf{end for}

\li $C^f \leftarrow  C^f + C_{temp}$
\end{codebox}
\vspace{-10pt}
\caption{A new version of ABFT for MM to facilitate the detection and correction of crash consistence in NVM.}
\label{fig:abft_pseudocode2}
\vspace{-15pt}
\end{figure}

The new algorithm solves the two problems in the original algorithm, because checksums in the result matrices ($C_s^{temp}$ and $C_{temp}$), once established and consistent in NVM, are not overwritten, hence can be used reliably to detect inconsistence of matrix data at any moment.
Furthermore, a set of temporal matrices enable easy recomputation and easy detection of inconsistence. %reduce recomputation cost.

The downside of the above algorithm extension is that we increase memory consumption. But with the deployment of NVM with much higher capacity than DRAM, we expect this problem is alleviated. Also, the memory consumption relines on a choice of $k$. A Smaller $k$ results in larger number of temporal matrices (more memory consumption) and smaller recomputation cost. We can manage memory consumption by selecting a good $k$ and exploiting the tradeoff between memory consumption and recomputation cost.

The above algorithm extension also increases the working set size which could cause extra cache misses and lose performance. However, we do not see big performance loss (no bigger than 8.2\%) in our evaluation. The reason is as follows. Given a large matrix size $n \times n$, the matrix multiplication based on the submatrix multiplication in the original code causes similar cache misses as the new algorithm, 
%because of fetching different submatricres for multiplication across iterations. 
because both of the original code and the new algorithm fetch different submatrices
for multiplication and save the results in either $C^f$ or $C_s$.
Those submatrices multiplications, which dominates the computation time,
have ``streaming-like'' memory access patterns --- we need to fetch submatrices one by one for multiplication. Such memory access patterns cause similar cache misses in the original code and the new algorithm. Furthermore, the regular memory access patterns in matrix multiplication allows prefetching to take effect and further alleviate the effects of cache misses.

\textbf{Performance evaluation.}
%Figure 5: runtime overhead (similar to CG)
%our method has much smaller overhead than checkpoint and intel pmem.
We evaluate the performance of our approach
%algorithm extension to ABFT for matrix multiplications 
from the perspective of recomputation cost and runtime performance.

\begin{figure}
\centering
\includegraphics[width=0.48\textwidth, height=0.15\textheight]{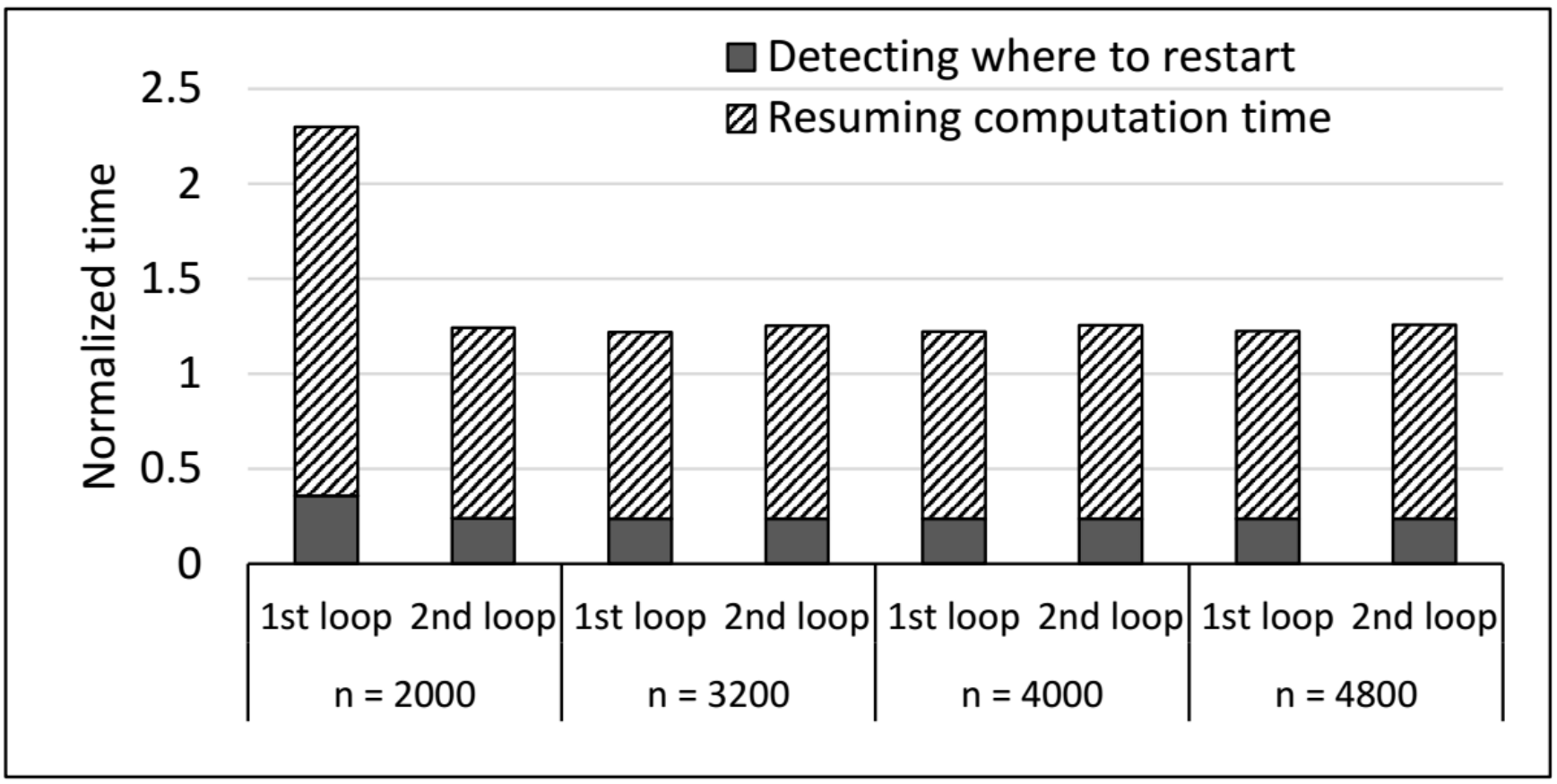}
\vspace{-10pt}
\caption{Recomputation cost (execution time) for ABFT-based matrix multiplication for two crash tests happened in the first and second loops. 
The $x$ axis is the matrix size. Rank $k=400$ in all tests.
The recomputation cost is normalized by the average execution time of one iteration in the first loop or the second loop in Figure~\ref{fig:abft_pseudocode2}.}
\label{fig:abft_recomp1}
\vspace{-20pt}
\end{figure}

Figure~\ref{fig:abft_recomp1} shows the recomputation cost on the heterogeneous NVM/DRAM system. 
%Similar to the analysis for CG, we break down the recomputation cost into two parts, the time to detect where to restart and the resumption time once where to restart is determined. 
We use four different input matrix sizes. For each matrix size, %we conduct random crash for five times with the crash emulator.
we do two crash tests.
One crash test triggers a crash at the end of the 4th iteration of the first loop in Figure~\ref{fig:abft_pseudocode2}. 
The second crash test triggers a crash at the end of the 4th iteration
of the second loop in Figure~\ref{fig:abft_pseudocode2}.
Hence, within Figure~\ref{fig:abft_recomp1},
we have two columns representing the recomputation cost
for the two crash tests for each matrix size.

The recomputation time for the first crash test is normalized by the average execution time of one iteration (i.e., submatrix multiplication) of the first loop in Figure~\ref{fig:abft_pseudocode2}.
The recomputation time for the second crash test is normalized by the average execution time of one iteration (i.e., submatrix addition) of the second loop in Figure~\ref{fig:abft_pseudocode2}.
Such normalization can quantify how many submatrix multiplications or additions are lost when a crash happens. 
Similar to the recomputation result for CG, Figure~\ref{fig:abft_recomp1} breaks down the recomputation cost 
into ``detecting where to restart'' and ``resuming computation time''.

\begin{comment}
For each crash test,  we start the recomputation based on the output data values in NVM. %the cache due to the cache emulator. 
According to the above result of cache simulator, when the crash happens at the end of iteration $i$, most of the data related to $C$ is already updated.
So we suppose there is one wrong column and estimate the recomputation time of different input size N. The time of verify checksum is very small compared to the total recomputation time (less than 5\% , input size N from 2000 to 5000), so it is not drawn in the figure.
\end{comment}
 
Figure~\ref{fig:abft_recomp1} shows that using different matrix sizes as input we lose different numbers of submatrix multiplication in the first crash test. With $n=2000$ as input, we lose about two submatrix multiplications, but with larger inputs, we lose only one submatrix multiplication. Using a larger input, our algorithm-based approach limits the recomputation to at most one submatrix multiplication because the submatrices are large and eliminate computation results from other iterations out of caches. 
For the second crash test, we always lose one submatrix addition, even if
we use a relatively small matrix (i.e., $n=2000$). The reason is that each iteration of the second loop has larger memory footprint than that of the first loop, which eliminates the results of submatrix addition from other iterations out of caches.

Note that during our crash tests, our algorithm-based approach cannot correct those inconsistent data in $C^{temp}_s$ and $C_{temp}$,
because there are too much inconsistent data in the same row or column, which are not correctable by the checksums.
Hence we claim in the above discussion that the submatrix multiplication or addition is lost. However, for some cases, it is possible that our algorithm-based approach can directly correct those inconsistence data based on checksums. 
In those cases, the submatrix multiplication or addition is not lost, and
the recomputation cost is even smaller.

\begin{figure*}[htp]
  \centering
  \subfigure[rank = 200]{\includegraphics[width=0.3\linewidth]{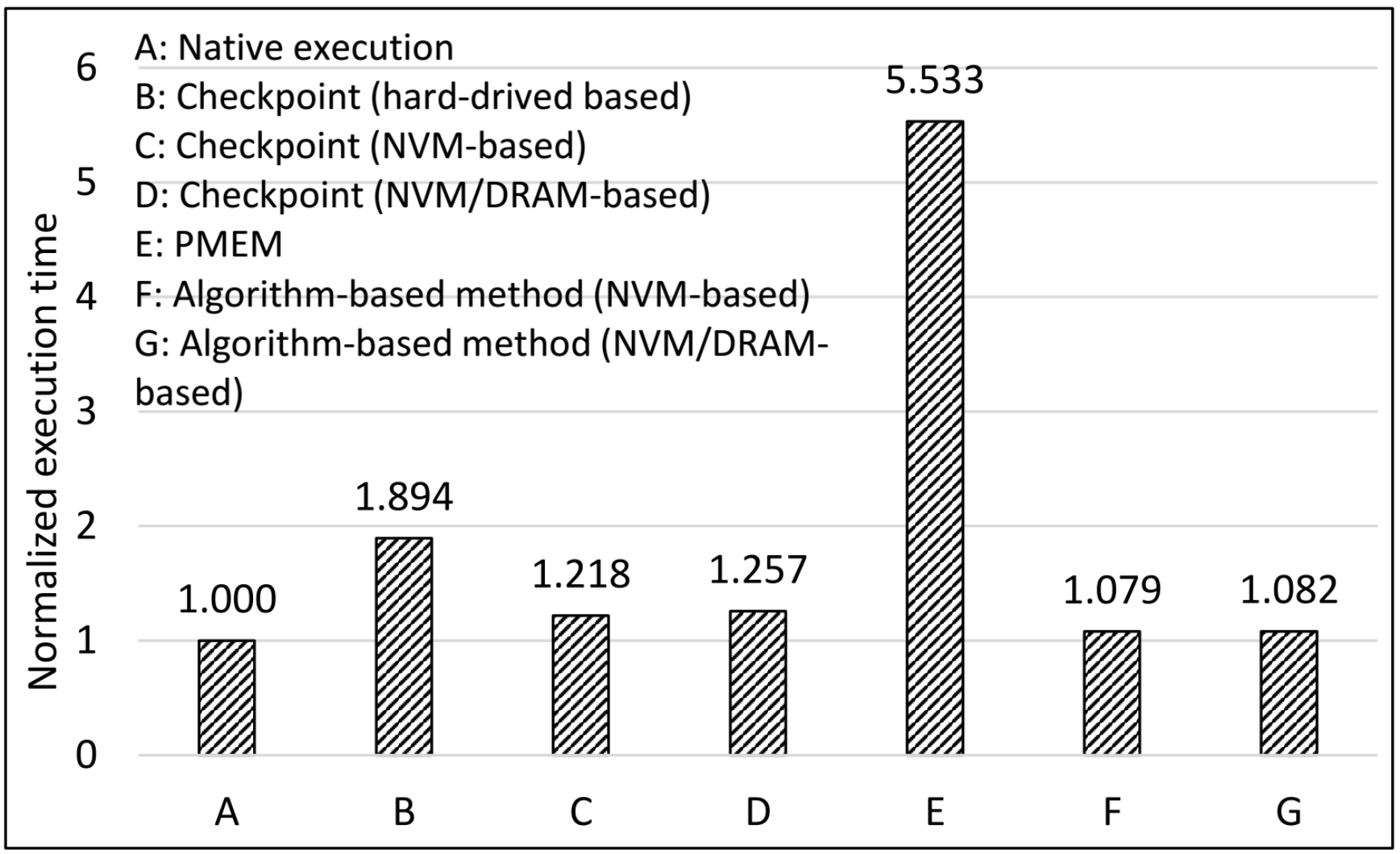}}
  \subfigure[rank = 400]{\includegraphics[width=0.3\linewidth]{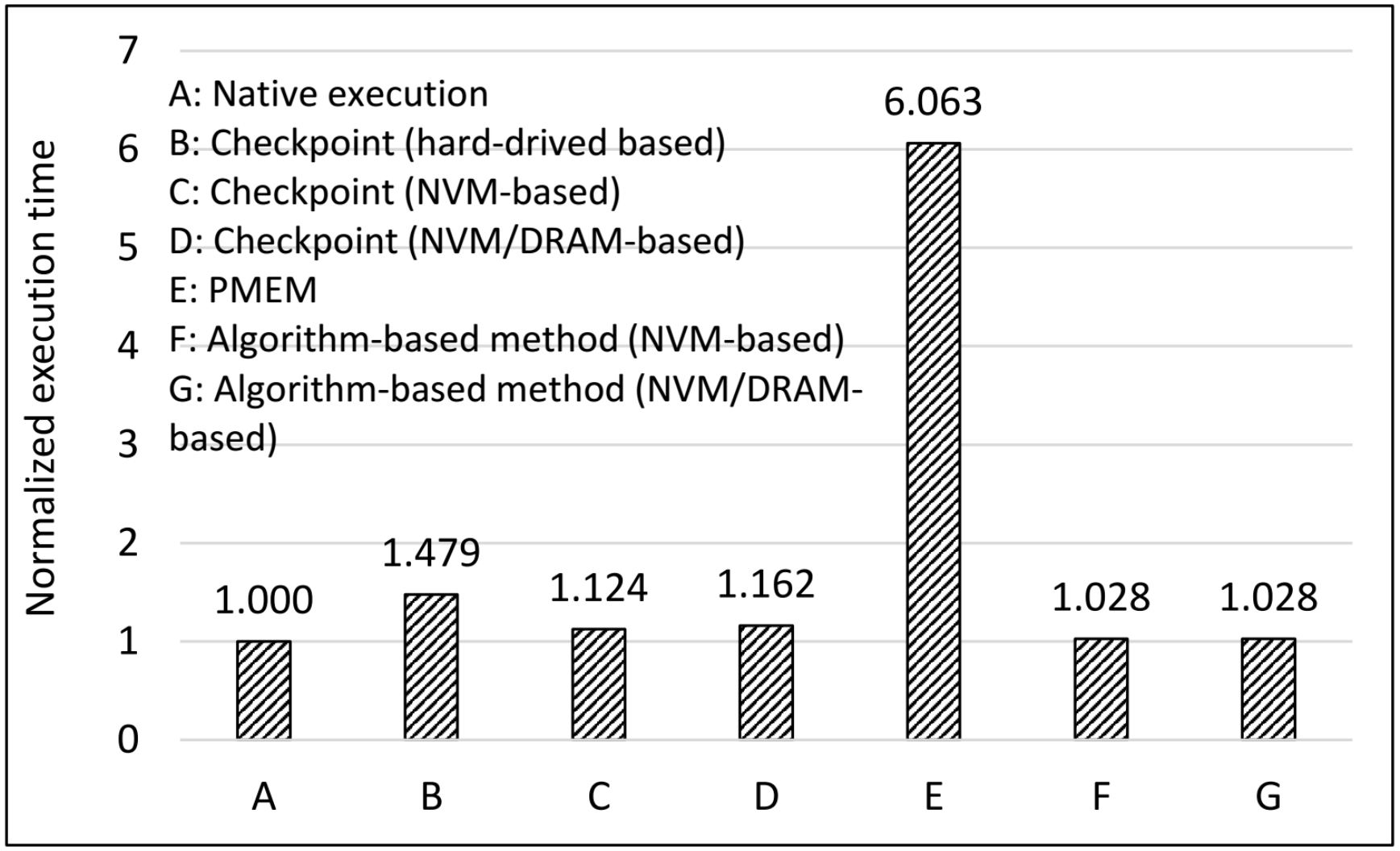}}
  \subfigure[rank = 1000]{\includegraphics[width=0.3\linewidth]{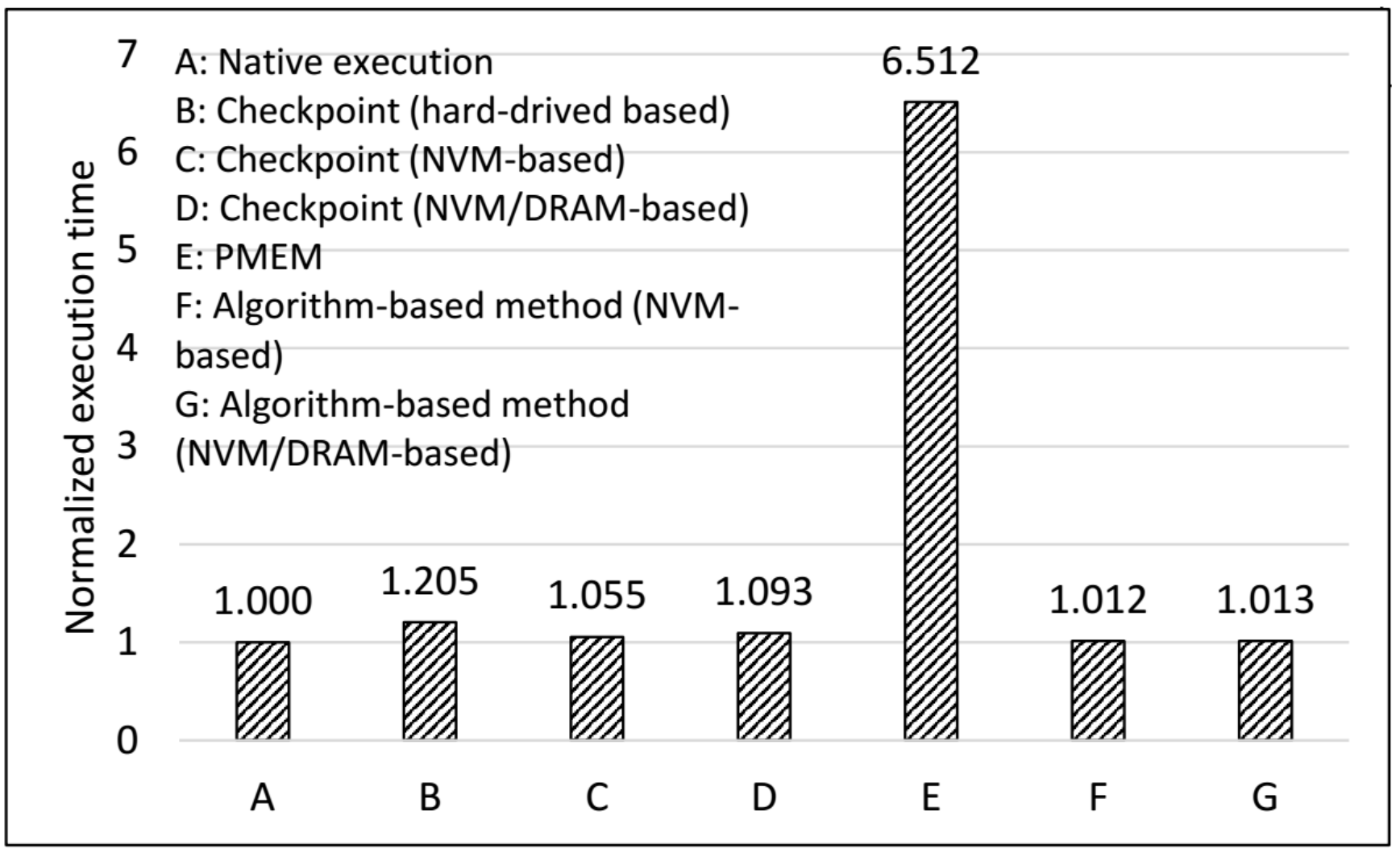}}
  \vspace{-10pt}
\caption{Runtime performance (execution time) with various checkpoint mechanisms and our algorithm extension for ABFT-based matrix multiplication. The matrix size is $n=8000$. Performance is normalized to the native execution with neither checkpoint nor our algorithm extension.}
\label{fig:abft_runtime} 
\vspace{-15pt}
\end{figure*}

We further study runtime performance. Similar to CG, we compare runtime performance with traditional checkpoint, Intel PMEM library~\cite{intel_nvm_lib}, and our approach. 
We use the matrix size as $n=8000$. 
Given such large matrix size, the recomputation cost
with our approach is limited to a submatrix multiplication
or a submatrix addition. %shown in Figure~\ref{fig:abft_recomp1}. 
Hence, we perform the traditional checkpoint at the end of each submatrix multiplication, such that the recomputation cost for the traditional checkpoint is a submatrix multiplication. 
%Because the recomputation cost with our approach is up to a submatrix multiplication, 
%we perform the traditional checkpoint at the end of each submatrix multiplication, such that the recomputation cost for the traditional checkpoint is also a submatrix multiplication.
For the PMEM library, each submatrix multiplication (the first loop of Figure~\ref{fig:abft_pseudocode2}) is a transaction and we enable transaction update on the submatrix multiplication result, which makes the recomputation cost also limited to a submatrix multiplication. Figure~\ref{fig:abft_runtime} shows the runtime performance.

The figure reveals that our algorithm-based approach has the smallest runtime overhead among all cases (no bigger than 8.2\% in all cases, depending on the rank size). Also, a larger rank size results in a smaller runtime overhead, because the algorithm does not need to frequently flush checksum cache blocks. With a large rank size (e.g., $rank=1000$), the runtime overhead is only 1.3\%.  
%(\textbf{Shuo TODO: what is the overhead?}). 

The runtime overhead of our algorithm-based approach is much smaller than the NVM-based checkpoint. 
For example, when $rank=200$, NVM-based checkpoint has at least 21.8\% overhead, while our approach has only 8.2\%.
This is due to the fact that our approach selectively flushes cache blocks of the checksums, instead of copying all data blocks of critical data objects. 
%Selectively flushing cache blocks is especially beneficial for a heterogeneous NVM/DRAM system with a relatively large DRAM cache.
%This small cache flushing overhead is especially pronounced in the heterogeneous NVM/DRAM system which includes flushing the DRAM cache. 

\begin{comment}
\begin{figure}
\centering
\includegraphics[width=0.48\textwidth]{figures/abft_recomp2.pdf}
\caption{Time of recomputation to the runtime of one iteration under different input size.}
\label{fig:abft_recomp2}
\end{figure}
\end{comment}

\textbf{Conclusions.} Treating inconsistent data as data corruption and using algorithm-based fault tolerance to detect and correct data inconsistence in NVM 
save expensive runtime overhead to establish crash consistence. Selectively flushing cache blocks to maintain crash consistence of checksums is the key.  
Given a large matrix size, our approach can effectively limit the recomputation cost to one submatrix multiplication or addition. 
%cache flushing operations. As a result, our approach has much smaller runtime overhead than the traditional approaches. 
%Furthermore, because of the error correction capability of ABFT, the recomputation cost can be much smaller than the traditional checkpoint-based approaches.

\subsection{Algorithm-Directed Crash Consistence for Monte Carlo Transport Simulation}
\label{sec:mc}

%Dong's comments:
%(1) For MC, our runtime performance will be similar to checkpoint and intel 
%pmem library; but our method can have smaller recomputation overhead 
%traditional checkpoint.
%Figure 9: when the crash happens, do we have a bid difference at the final results, if we immediately restart?
%Story 1: Monte Carlo shouldn't have recomputation;
%Story 2: We will show that a data object could still be lost a lot if we don't clflush timely.

Monte Carlo method (MC) has been applied to a broad range of scientific simulations, such as nuclear reactor physics and medical dosimetry. In essence, MC employs repeated random sampling to obtain numerical results and solve problems that are deterministic in principle. Given MC simplicity, MC provides significant advantages compared to deterministic methods.
Leveraging MC's random nature, we explore how to use MC algorithm knowledge to 
build crash consistence in NVM without losing scientific simulation accuracy.

%calculate the distribution and generation rates of neutrons within a nuclear reactor.
We focus on a specific MC benchmark, XSBench. This benchmark models
the calculation of macroscopic neutron ``cross sections''~\cite{xsbench2:john} within a nuclear reactor, which is the most computationally intensive part of a typical MC transport algorithm~\cite{xsbench:john}. Listing~\ref{fig:xsbench_pseudocode} shows the major computation of XSBench.

XSBench has two large, read-only data arrays, which are a nuclide grid and an energy grid. The two arrays account for most of the memory footprint of XSBench. 
%A large data structure that holds cross section data points for many discrete energy levels. This data structure is read-only.
XSBench has a main computation loop, and each iteration of the loop performs a 
lookup of some data (particularly ``cross section'' data) from the nuclide grid
with the assist of searching the energy grid. 
The lookup result of each iteration accumulates to an array (particularly, macro\_xs\_vector) with five elements (Line 7 in Figure~\ref{fig:xsbench_pseudocode}). Each element of the array macro\_xs\_vector is the value of a macroscopic cross section.
Those five values correspond to five different particle interaction types in the nuclear reactor.
In each iteration, the lookup happens based on two randomly chosen inputs (particularly, neutron energy and material, shown in Line 2 in Figure~\ref{fig:xsbench_pseudocode}).

XSBench is only a benchmark for performance study, hence its result (particularly macro\_xs\_vector) does not have sufficient physical meaning. 
From one run to another, the result can be different due to the random nature of the benchmark. It is difficult to know if the benchmark result remains correct for our crash consistence evaluation. Based on the domain knowledge, we slightly extend the benchmark such that the benchmark result has physical meaning.
In particular, at the end of each iteration, 
we apply a cumulative distribution function (CDF) to the five elements
of macro\_xs\_vector, and then normalize the CDF result by the largest element. Then we generate a uniformly distributed random number $x$ ( $0<x<1$). %following uniform distribution.
%Mersenne-Twister algorithm
This random number represents a computation result in a full-featured simulation of the nuclear reactor.
Based on the random number, we find which interaction type (i.e., which element of macro\_xs\_vector) should be chosen based on the normalized CDF result. 

For example, suppose macro\_xs\_vector = \{0.9, 0.1, 0.3, 0.6, 0.05\} in one iteration.
We create a CDF of this macro\_xs\_vector, which is \{0.9, 1.0, 1.3, 1.9, 1.95\}. Then we normalize the CDF result by the largest element of macro\_xs\_vecotr (i.e., 1.95).
The normalization result is \{0.462, 0.513, 0.667, 0.974, 1.0\}. 
Then, we choose a random number, for example, 0.65. According to the normalization result, 0.65 falls between the second (0.513) and third elements (0.667),
which means the second interaction type is chosen.

We perform the above process for each iteration of the XSBench loop, and introduce five counters to count how many times each interaction type is chosen for all iterations.
%(or lookups).
Given the sufficient number of lookups (i.e., iterations of the main computation loop), the number of times an interaction type is chosen is roughly the same for all interaction types. This method gives us a deterministic and meaningful way to quantify the validness of the benchmark result.

\begin{comment}
\begin{lstlisting}[language=c++,caption=Pseudo code for XSBench, label=listing:xsbench_pseudocode, linewidth=9cm]
for (i=0; i < lookups; i++) {
  Generates two randomly sampled inputs (neutron energy and material);
  Binary search on the energy grid;
   for each nuclide in the input material{
     look up two bounding nuclide grid points from the nuclide grid;
     interpolate the two points to give microscopic cross sections;
     microscopic cross sections accumulate into macroscopic cross section;
   }
}
\end{lstlisting}
\end{comment}

\begin{figure}
\centering
\small
\begin{codebox}
%\Procname{$\proc{XSBench macroscopic cross section lookup}$}
\li \textbf{for} (i = 0; i $<$ total number of lookups; i++)
\li \quad  Generates two randomly sampled inputs \\
     \quad (neutron energy and material);
\li \quad  Binary search on the energy grid;
\li \quad \textbf{for} each nuclide in the input material
\li \quad \quad Look up two bounding nuclide grid points \\
    \quad \quad from the nuclide grid;
\li \quad \quad Interpolate the two points to give microscopic \\       
    \quad \quad cross sections;
\li \quad \quad Microscopic cross sections accumulate into \\
    \quad \quad macroscopic cross section (i.e., macro\_xs\_vector);
\li \quad \textbf{end for}
\li \textbf{end for}
\end{codebox}
\vspace{-15pt}
\caption{Pseudo code for XSBench.}
\label{fig:xsbench_pseudocode}
\vspace{-20pt}
\end{figure}

\textbf{Algorithm extension.}
Because of the random nature of XSBench, we expect that directly restarting
from remaining data in NVM after a crash does not result in incorrect results, assuming that macro\_xs\_vector and the two arrays (nuclide grid and energy grid) in NVM are accessible after the crash. In particular, at every iteration, we flush the cache block containing the loop index variable $i$, such that we can know which iteration the crash happens. After the crash happens, we restart from the beginning of the iteration $i$. Except flushing the single cache block, we do not use any mechanism to establish data consistence and validness in NVM during the execution. %of XSBench.

Using the above approach, we may lose the binary search result (Line 3 in Figure~\ref{fig:xsbench_pseudocode}), the lookup result in the nuclide grid (Line 5 in Figure~\ref{fig:xsbench_pseudocode}), the interpolate result (Line 6 in Figure~\ref{fig:xsbench_pseudocode}), and the accumulation result (Line 7 in Figure~\ref{fig:xsbench_pseudocode}).
However, we expect the impact of those results losses is limited, because the binary search and the lookup in the nuclide grid need to load the energy grid and the nuclide grid into caches. Those two grids are two large arrays, and 
Loading them can evict the results of most of the previous iterations out of caches, and implicitly enable data consistence.
We may only lose the results of a few iterations, and those results will not be accumulated into macro\_xs\_vector (Line 7 in Figure~\ref{fig:xsbench_pseudocode}).
However, losing a few iterations of the results may not impact
the accuracy of counting the number of times each interaction type is chosen, because XSBench uses a sampling-based approach and takes a number of samples (i.e., the number of lookups shown in Line 1).
As the number of samples is large, losing a few samples is not expected to impact the counting accuracy.
%Also, we expect that losing \textbf{a few} iterations of the results of microscopic cross sections without accumulating them into macro\_xs\_vector does not impact the final result of macro\_xs\_vector and counting the number of interaction type, because of the sampling nature of XSBench.

To verify the above basic idea, we run XSBench with an input problem of 34 fuel nuclides in a Hoogenboom-Martin reactor model. 
With such input problem, the energy grid and nuclide grid take about 246MB memory.
There are $1.5 \times 10^7$ lookups in the main computation loop.
We use our crash simulator to run the benchmark and trigger a crash when the benchmark is in the $1.5 \times 10^6$th lookup (10\% of all lookups). 

\begin{figure}
\centering
\includegraphics[width=0.4\textwidth, height=0.12\textheight]{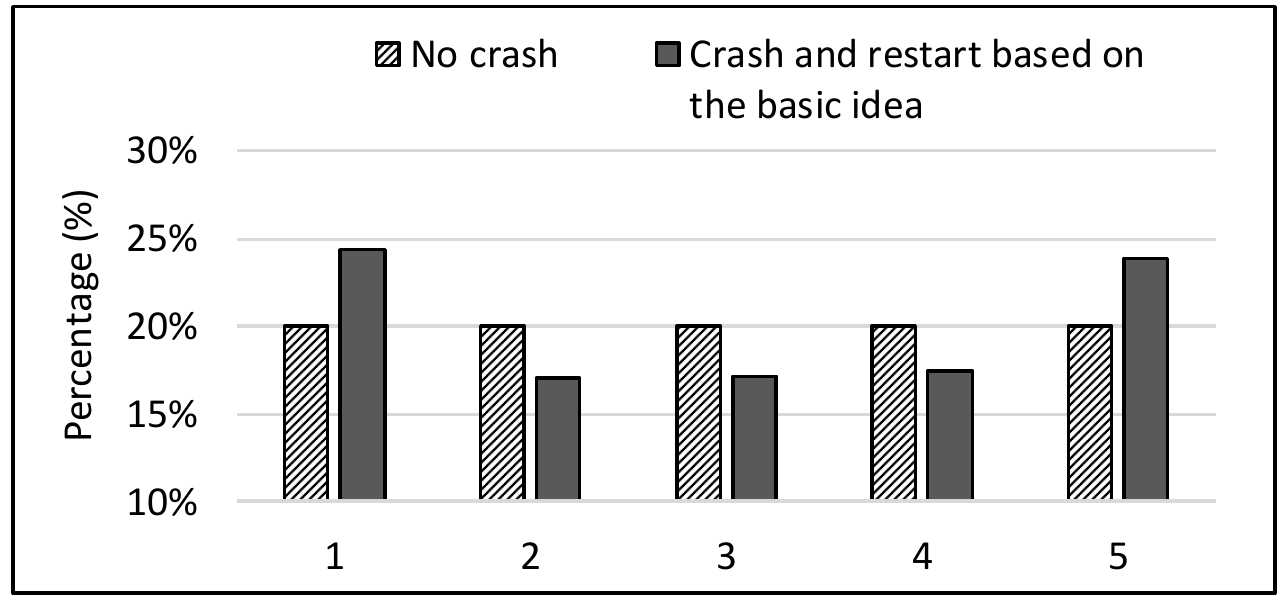}
%\caption{After crash, if we missed part of the elements of xs\_vector.}
\vspace{-10pt}
\caption{Comparing XSBench results of two cases (one case without crash and the other case with crash and restart based on the basic idea).}
\label{fig:xsbench_exp2}
\vspace{-10pt}
\end{figure}

Figure~\ref{fig:xsbench_exp2} shows how many times each interaction type
is counted for two tests.  In one test, we do not have crash (labeled as ``No crash''); in the other test, we have the crash but immediately restart based on the above basic idea (labeled as ``Crash and restart based on the basic idea'').
The numbers of times counted for the five interaction types are normalized by the total number of lookups and shown as percentage in the $y$ axis.
These two tests use the same randomly sampled inputs (Line 2 in Figure~\ref{fig:xsbench_pseudocode}) for each lookup, such that we enable a fair comparison of the XSBench results of the two tests.
%There are five interaction types shown in the $x$ axis.

From Figure~\ref{fig:xsbench_exp2}, we notice that the five interaction types in the case of no crash have almost the same counting result. 
%the case without crash has evenly distributed counting results between the five interaction types. This is expected, given the random and sampling nature of XSBench. %and its result. 
However, the case with crash and restart based on the basic idea has obviously different counting results for the five interaction types. For example, there is 8\% difference in the counting result between the interaction types 1 and 2.

To investigate the reason why there is such result difference between the two cases,
we examine the five counters in NVM at the crash trigger point,
and compare the values of those counters in the two cases.
We found that the values of the five counters in the two cases are very different.
For the crash case, the counting results from a number of iterations before the crash trigger point are still in caches. They are not evicted out of caches as expected. %consistent with those in NVM.
For the crash case, the counting results are not consistent between caches and NVM. 

Further investigation reveals that although the two grids are two large arrays, 
binary search on the energy grid (Line 3 in Figure~\ref{fig:xsbench_pseudocode}) and 
lookup operation in the nuclide grid (Line 5 in Figure~\ref{fig:xsbench_pseudocode})
do not necessarily access the whole grids, and hence the five counters
and macro\_xs\_vector are not evicted out of caches in many iterations as expected. Also, the five counters and macro\_xs\_vector are frequently updated from one iteration to another.
Such frequent updates keep those variables in caches and make them inconsistent between caches and NVM. Hence, when a crash happens, we lose the results of many iterations.

To address the above data inconsistence problem, we can flush the five counters and macro\_xs\_vector whenever there is any update happened to them at every iteration. However, such frequent cache flushing causes 16\% performance loss.
Hence, we flush caches lines every $n$ iterations ($n=$ 0.01\% of total number of lookups), shown in Figure~\ref{fig:xsbench_pseudocode2}. 
%shows the pseudo code for the new XSBench. 

\begin{figure}[tpb]
\centering
\small
\begin{codebox}
%\Procname{$\proc{XSBench macroscopic cross section lookup}$}
%\li \textbf{for} (i = 0; i $<$ total number of lookups \\ {\color {red} (15,000,000 lookups)}; i++)
\li \textbf{for} (i = 0; i $<$ total number of lookups; i++)
\li \quad  Generates two randomly sampled inputs \\
     \quad (neutron energy and material);
\li \quad  Binary search on the energy grid;
\li \quad \textbf{for} each nuclide in the input material
\li \quad \quad Look up two bounding nuclide grid points \\
    \quad \quad from the nuclide grid;
\li \quad \quad Interpolate the two points to give microscopic \\       
    \quad \quad cross sections;
\li \quad \quad Microscopic cross sections accumulate into \\
    \quad \quad macroscopic cross section (i.e., macro\_xs\_vector);
%\li \quad \quad {\color {red} if (every 1500 lookups 0.01\% of total number of lookups)}
\li \quad \quad {\color {red} if (every 0.01\% of total number of lookups)}
\li \quad \quad \quad {\color {red} flushing macro\_xs\_vector, the five counters and $i$;} 
\li \quad \textbf{end for}
\li \textbf{end for}
\end{codebox}
\vspace{-10pt}
\caption{Selectively flush cache blocks for XSBench based on the algorithm knowledge. Our extension to XSBench is highlighted in red.}
\label{fig:xsbench_pseudocode2}
\vspace{-15pt}
\end{figure}

\begin{figure}[btp]
\centering
\includegraphics[width=0.45\textwidth, height=0.15\textheight]{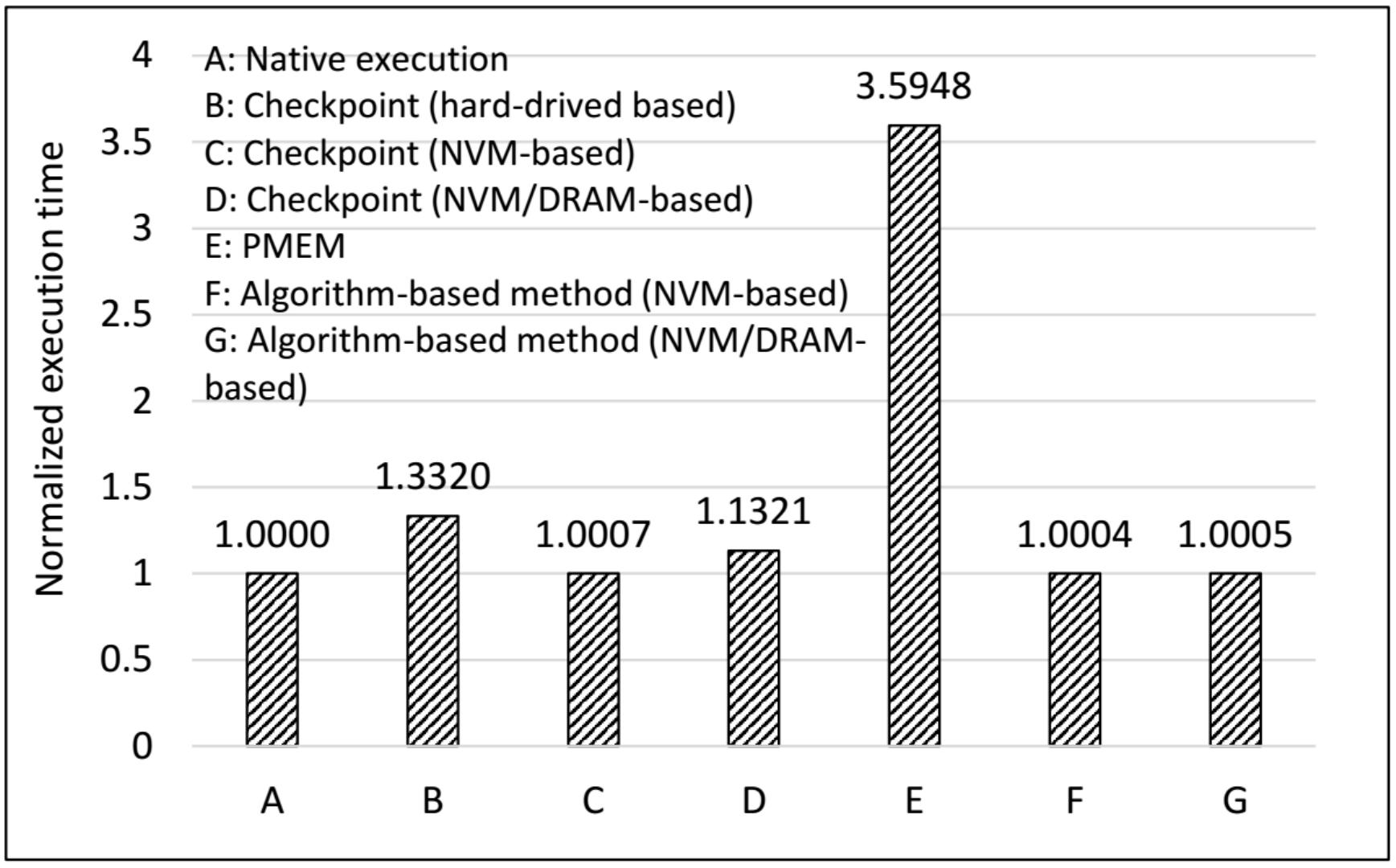}
\vspace{-10pt}
\caption{Runtime performance (execution time) with various checkpoint mechanisms and our algorithm-based approach.}
\label{fig:xsbench_runtime}
\vspace{-20pt}
\end{figure}

How often we should flush cache blocks is a challenging problem.
For XSBench, if the sizes of the two grids are big and a large portion of the two grids are touched at each iteration, then the five counters and macro\_xs\_vector can be evicted out of
caches automatically and frequently, and we do not have to frequently flush cache blocks.
However, due to the randomness of sampled inputs (i.e., neutron energy and material at Line 2 in Figure~\ref{fig:xsbench_pseudocode2}),
it is difficult to quantify the memory footprint size of each iteration. %for the binary search and lookup on the two grids. 
We empirically determine the frequency as every 0.01\% of total number of lookups in our tests. Such frequency of cache flushing has ignorable performance overhead shown in Figure~\ref{fig:xsbench_runtime}. Also, using such frequency for flushing cache blocks, we bound the result loss when a crash happens by 0.01\% of total number of iterations, which is small.

\begin{figure}[btp]
\centering
\includegraphics[width=0.4\textwidth, height=0.12\textheight]{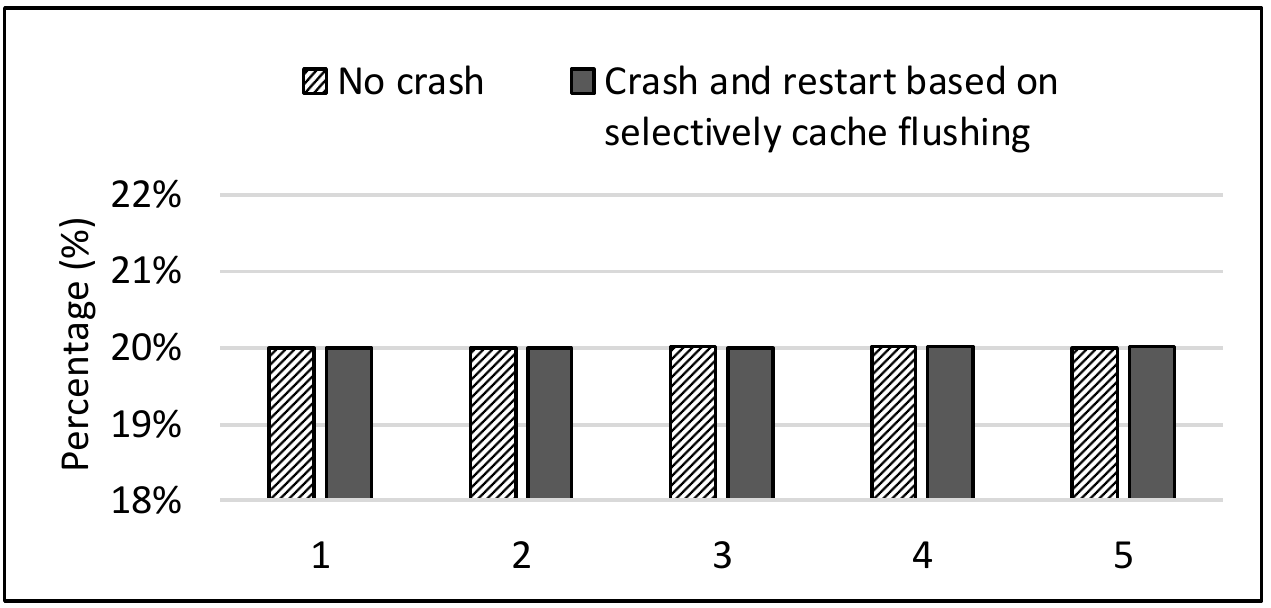}
\vspace{-10pt}
\caption{Comparing XSBench results of two cases (one case without crash and the other case with crash and restart based on selectively cache line flushing).}
\label{fig:xsbench_exp1}
\vspace{-10pt}
\end{figure}

\textbf{Performance evaluation.}
%Figure 8: runtime overhead (similar to CG).
%The five cases should have the similar performance.
We first compare the result correctness between our approach and no-crash case. We trigger the crash at the same point as the one in Figure~\ref{fig:xsbench_exp2}, i.e., the $1.5 \times 10^6$th lookup (10\% of all lookups). Figure~\ref{fig:xsbench_exp1} shows the result. With our approach, the number of times an interaction type is chosen is almost the same for all interaction types, which is the same result as the one with no crash.

%we compare runtime performance with traditional checkpoint, Intel PMEM library~\cite{intel_nvm_lib}, and our approach.  Since xs\_vector is the only data object that needs to be persistent, we checkpoint it after each lookup. For the PMEM library, we make each lookup a transaction and enable transactional update on xs\_vector. Figure~\ref{fig:xsbench_runtime} shows the results with the input grid size as large (i.e., the 5678MB grid). 

We further compare the performance of the seven cases. 
For those cases with checkpoint, we checkpoint macro\_xs\_vector and five counters at every 0.01\% of
total number of iterations. This checkpoint frequency is the same as that in our algorithm.
%Such checkpoint frequency is the same as that for cache block flushing.
Figure~\ref{fig:xsbench_runtime} shows the results.
Our selective cache block flushing (labeled as ``algorithm-based approach'') has ignorable overhead (at most 0.05\%).
NVM-based checkpoint based on the NVM-only system %having the same performance as DRAM 
also has ignorable overhead. %performance loss. 
However, when we use the NVM/DRAM system, checkpoint overhead is as large as 13\%, much larger than our runtime overhead.

%In this figure, we do the checkpoint every 0.01\% of total lookups (do checkpoint every 1500 lookups). The figure shows that there is no big performance difference across all cases, because xs\_vector is rather small and frequently saving it into the persistent storage (either hard drive or NVM) does not cause performance loss. Meanwhile, the checkpoint (hard-drived based) have more than 200\% overhead than the native execution. 

\textbf{Conclusions.}
%Monte Carlo-based simulation can tolerate data inconsistence due to its random nature. However, we reveal that we still need to enable data persistency for critical data objects. Those data objects are small, hence flushing them out of caches has ignorable runtime overhead. 
To ensure result correctness, MC simulation must flush a few cache blocks to enable crash consistence. %for a few critical data objects.
Different from CG which has large data objects frequently evicting critical data objects out of caches, XSBench with large data objects may have small memory footprint at each iteration and cannot evict critical data objects. This is due to the random nature of MC simulation.

\section{Related Work}
\label{sec:related}
\begin{comment}
We discuss related research efforts that 
implement crash consistence in NVM.
%seek to leverage the non-volatility of NVM 
%Those research efforts explore how to enforce write-ordering to build persistent memory. 
We also discuss the recent progress of algorithm-based fault tolerance in HPC in this section.
\end{comment}

%\textbf{Persistent memory.}
\textbf{Crash consistence in NVM.}
Leveraging persistent extensions from ISA (e.g., {\fontfamily{qcr}\selectfont CLFLUSH}), some work introduces certain program constructs to enable 
crash consistence in NVM.
%%persistent memory. 
%Comparing with the work discussed in the last paragraph, this work needs much less hardware support. 
Mnemosyne~\cite{mnemosyne_asplos11}, Intel NVM library~\cite{intel_nvm_lib, usenix13:rudoff}, NV-heaps~\cite{nv-heaps_asplos11}, and REWIND~\cite{vldb_endow15:chatzistergiou} provide transaction systems optimized for NVM. 
NVL-C~\cite{hpdc16:denny} introduces flexible directives and runtime checks that guard against failures that corrupt data consistence. %memory persistency.
SCMFS~\cite{sc11:wu} provides a PM-optimized file system based on the persistent extensions from ISA.
Atlas~\cite{oopsla14:dhruva} uses those extensions for lock-based code. To use the existing efforts for HPC applications, we may have to make extensive changes to applications or operating systems. The application can suffer from large runtime overhead because of frequent runtime checking or data logging.
Our evaluation with the Intel NVM library shows such large overhead.

Some work introduces persistent cache, such that stores become durable as they execute~\cite{micro13:zhao, vldb_endow14:wang, asplos12:dushyanth}. Those existing efforts eliminate the necessity of any cache flushing operation, by not caching NVM accesses, or by ensuring that a batter backup is available to flush the contents of caches to NVM upon power failure. However, those existing efforts need extensive hardware modification. It is not clear if integrating NVM into processors has any manufacturing challenges.

Some work divides program execution into epochs. In the epoch, stores may
persistent concurrently by flushing cache lines or bypassing caches.
%BPFS~\cite{sosp09:condit} is one of the pioneer work that explores the idea of epochs for NVM.  BPFS implements epochs by tightly coupling with cache management. In particular, BPFS tags all cache blocks with an epoch ID on every store and modifies the cache replacement policy to write epochs back to NVM in order. 
Pelley et al.~\cite{Pelley:isca14} introduce a couple of variation of epoch,
and demonstrate potential performance improvement because of a relaxation of
inter-thread persist dependencies. 
Joshi et al.~\cite{micro15:joshi} propose a buffered epoch persistency by defining efficient persist barriers.
Delegated ordering~\cite{micro16:kolli} decouples cache management from the path persistent writes take to memory to allow concurrent writes within the same epoch to improve performance.
Those existing efforts can be complementary to our work to improve the performance of cache flushing (especially for algorithm-directed crash consistence based on ABFT for matrix multiplication).

\textbf{Algorithm-based program optimization.}
%HPC applications heavily rely on numerical algorithms. 
Leveraging algorithm knowledge is an effective approach
to improve performance~\cite{ipdps14:faverge, Lam:asplos91, Williams:sc07}, application fault tolerance~\cite{Chen:2013ie, ft_lu_hpdc13, ftfactor_ppopp12, jcs13:wu, abft_ecc:SC13}, 
and energy efficiency~\cite{Dorrance:fpga14, Garcia:lcpc13}. 
Different from the existing efforts, this paper uses algorithm knowledge to achieve crash consistence in NVM.
This is a fundamentally new approach to explore the usage of algorithm knowledge. %for future HPC.

  %1
\section{Conclusions}
\label{sec:conclusions}
Leveraging the emerging NVM to establish crash consistence is a promising while challenging approach to enable resilient HPC. HPC has high requirement for performance. We must minimize runtime overhead when building crash consistence. This paper introduces a fundamentally new methodology to do so. Based on the algorithm knowledge and comprehensive performance evaluation, we show that we can significantly reduce runtime overhead while enabling detectable crash consistence for future NVM-based HPC.
%comparing with the traditional checkpoint and a recent Intel library that implements crash consistence in NVM. Using the algorithm knowledge, we enable detectable crash consistence in NVM, and provide a solution to replace checkpoint for future NVM-based HPC.    %0.5page


\bibliographystyle{IEEEtran}
\bibliography{li}

\end{document}